\documentclass[journal]{IEEEtran}
\usepackage{xcolor,soul,framed} 

\colorlet{shadecolor}{yellow}
\usepackage[pdftex]{graphicx}
\graphicspath{{../pdf/}{../jpeg/}}
\DeclareGraphicsExtensions{.pdf,.jpeg,.png}

\usepackage{array}
\usepackage{mdwmath}
\usepackage{mdwtab}
\usepackage{eqparbox}
\usepackage{cite}
\usepackage{amsmath,amssymb,amsfonts,dsfont}
\usepackage{algorithmic}
\usepackage[pdftex]{graphicx}
\usepackage{textcomp}
\usepackage{caption}
\usepackage{color,soul}
\usepackage{url}
\usepackage{subfigure}
\usepackage{epstopdf}
\usepackage{multirow}
\usepackage{bm}
\usepackage{ulem}
\begin{document}
\bstctlcite{IEEEexample:BSTcontrol}
    \title{A Low-cost, High-impact Node Injection Approach for Attacking Social Network Alignment}
  \author{Shuyu Jiang, Yunxiang Qiu, Xian Mo, Rui Tang, and Wei Wang

\thanks{This work was supported by the National Natural Science Foundation of China under Grant Nos. 62202320, 62306157; the Fundamental Research Funds for the Central Universities under Grant No. 2023SCU12126. (Corresponding author:Rui Tang. e-mail: tangrscu@scu.edu.cn)}
  \thanks{Shuyu Jiang is with School of Cyber Science and Engineering, Sichuan University, Chengdu 610065, China.}
  \thanks{Yunxiang Qiu is with School of Cyber Science and Engineering, Sichuan University, Chengdu 610065, China; the Key Laboratory of Data Protection and Intelligent Management, Ministry of Education, Sichuan University, Chengdu 610065, China; and Huaxin Consulting and Designing Institute Co. Ltd, Hangzhou, 310052, China.}
  \thanks{Xian Mo is with the School of Information Engineering, Ningxia University, Yinchuan, 750021, China.}
  \thanks{Rui Tang is with the School of Cyber Science and Engineering and the Key Laboratory of Data Protection and Intelligent Management, Ministry of Education, Sichuan University, Chengdu 610065, China.}%
  \thanks{Wei Wang is with the School of Public Health, Chongqing Medical University, Chongqing, 401331, China.}
  }

\markboth{IEEE
}{IEEE}

\maketitle

\begin{abstract}
Social network alignment (SNA) holds significant importance for various downstream applications, prompting numerous professionals to develop and share SNA tools. Unfortunately, these tools can be exploited by malicious actors to integrate sensitive user information, posing cybersecurity risks. While many researchers have explored attacking SNA (ASNA) through a network modification attack way, practical feasibility remains a challenge. This paper introduces a novel approach, the node injection attack. To overcome the problem of modeling and solving within a limited time and balancing costs and benefits, we propose a low-cost, high-impact \textbf{n}ode \textbf{i}njection \textbf{a}ttack via \textbf{d}ynamic \textbf{p}rogramming (DPNIA) framework. DPNIA models ASNA as a problem of maximizing the number of confirmed incorrect correspondent node pairs who have a greater similarity scores than the pairs between existing nodes, making ASNA solvable. Meanwhile, it employs a cross-network evaluation method to identify node vulnerability, facilitating a progressive attack from easy to difficult. Additionally, it utilizes an optimal injection strategy searching method, based on dynamic programming, to determine which links should be added between injected nodes and existing nodes, thereby achieving a high impact for attack effectiveness at a low cost. Experiments on four real-world datasets consistently demonstrate that DPNIA consistently and significantly outperforms various attack baselines.
\end{abstract}


\IEEEpeerreviewmaketitle
	
\section{Introduction}
Social network alignment (SNA) aims to identify whether the accounts in different social network platforms belong to the same users~\cite{ShuKai2017,cui2019survey,xu2023sinkhorn}. Due to the fact that billions of users worldwide simultaneously use multiple social network platforms every day, SNA has become a very popular research area in recent years. It is an fundamental task for many downstream applications such as e-commerce recommendation~\cite{KongXiangnan2013}, information diffusing analysis~\cite{chen2021adversarial}, user identity verification~\cite{liu2023wl}, and cross-network user profiling~\cite{gao2022reborn}. Researchers from both the industry and academia continuously enhance the capabilities of SNA algorithms, models, and tools from various perspectives, including accuracy~\cite{liu2023wl,ding2022user,tang2023interlayer}, efficiency~\cite{WangYongqing2019-www,tang2022interlayer,zhang2019attributed}, scalability~\cite{zeng2023parrot,trung2020adaptive}, etc.

Unfortunately, despite the advantages it brings to the various applications, SNA has also posed significant challenges in terms of cybersecurity risks. Malicious actors can utilize SNA tools to integrate sensitive information about users, such as occupation, age, address, email, hobbies, friends, and daily commute patterns, scattered across different social platforms. Numerous codes and tools related to SNA are available on the open-source sites. If someone input keywords like "social network alignment" or other related terms such as "user identity linkage", "anchor link prediction", and "graph matching" on GitHub, a well-known code sharing platform, they will find large amount of related repositories. These repositories encompass alignment approaches for various social network platforms, such as Facebook, Twitter, Instagram, and numerous others. Since malicious actors also have the opportunity to download the repositories, users of these platforms are exposed to the risks of sensitive privacy breaches. Subsequently, they and their families may fall victim to scams or manipulation. It is urgent for us to investigate methods for attacking social network alignment (ASNA) to efficiently and effectively safeguard users from such threats.

Previous studies of ASNA usually follow the setting of network modification attack (NMA) which perturb the network structure by manipulating the links between existing nodes. The basic operations involve removing existing links, switching existing links, or adding new links between existing nodes. For instance, Zhang et al.~\cite{zhang2020adversarial} pushed the nodes to dense regions by adding and removing links of the existing nodes, therefore degrade the quality of network alignment. Tang et al.~\cite{TANG2022109095} attacked the SNA methods by searching the most important links for correct aligning and removing them. Refs.~\cite{shao2023toak,zhang2021adversarial,shen2014defending} have also conducted relevant explorations. To fulfill the attack, all of these studies need to manipulate plenty of existing nodes. However, only the users of the accounts represented by the nodes on the platform have the authority to add or remove social links from them. The attacker must either gain the trust of a significant number of social network users and persuade them to carry out the required removing or addition operations, or take control of these accounts through hacking methods. Obviously, achieving the former is highly improbable, while individuals resorting to the latter will face legal consequences. Finding a more practical approach to perturb the network structure is crucial.

This paper introduces a novel node injection attack (NIA) approach for ASNA, which injects fake accounts into social network platforms and adds links between those nodes to carefully selected existing nodes. Because of the open nature of social network platforms, numerous users enjoy making new friends for friendship and readily accept friend requests from strangers~\cite{sun2020adversarial}. Therefore, such way is easier and more feasible in real scenarios than the way of NMA. It faces the following challenges: (i) How to model and solve the problem of adding links between injected nodes and existing nodes to effectively degrade the accuracy of SNA. As we known, performing attacks on network data is more difficult than performing attacks on image or text, because of the discrete and non-differential nature of the adjacency matrix of network data~\cite{wang2022cluster}. (ii) How to balancing the costs and benefits. Injecting fake nodes into social network platforms and adding links to existing nodes necessitate the utilization of various resources. Attackers would like to achieve significant attack effectiveness at a relatively low cost. (iii) How to prevent the NIA from being noticed by the defender. In the real-world scenarios, the attack and defense probable occur simultaneously~\cite{li2019attack}. The defender teams of social network platforms might kill off the injected nodes. (iv) How to find the optimal way of establishing links within a limited time. Unlike existing NIA studies on single-network tasks, such as attacking node classification via node injection~\cite{sun2020adversarial,wang2022cluster,zou2021tdgia,tao2021single}, ASNA needs to comparing the impact of established links on all of the pairings of nodes across multiple networks, leading to an exponential increase in time complexity.

To tackle the above problem, we propose a low-cost, high-impact \textbf{n}ode \textbf{i}njection \textbf{a}ttack via \textbf{d}ynamic \textbf{p}rogramming (DPNIA) framework for ASNA. It employs the designed cross-network evaluation method to identify node vulnerability, facilitating a progression of attacks from easy to difficult. Additionally, it utilizes the developed optimal injection strategy searching method, based on dynamic programming, to determine links to be added, thereby achieving a high impact for attack effectiveness within a low cost. The contributions of this paper are as follows:
\begin{itemize}
\item We propose a low-cost, high-impact DPNIA framework. In this framework, ASNA is modeled as a problem of maximizing the number of confirmed incorrect correspondent node pairs who have a greater similarity score than the node pairs between existing nodes, making ASNA solvable. To the best of our knowledge, this is the first exploration of NIA for ASNA.

\item We design a cross-network evaluation method to identify node vulnerability through theoretical analysis of similarity scores. Consequently, the attack can be implemented progressively from easy to difficult and our focus is solely on ensuring that the similarity scores of confirmed incorrect pairs surpass those of any other pairs via node injection. This eliminates the necessity for repeated comparisons of ASNA's impact on all node pairs across different social networks after injection, facilitating the problem's resolution within a constrained time.

\item To minimize costs and avoid detection by defenders, we develop an optimal injection strategy searching method based on dynamic programming to determine which links should be added between injected nodes and existing nodes, thereby achieving a high impact for attack effectiveness at a low cost. It seeks the minimum budget required for a successful attack on a single node through cross-network evaluation, and then utilizes dynamic programming to maximize the number of nodes that can be influenced with each budget allocation.

\item We conduct experiments on several typical real-world social network datasets, targeting multiple neighborhood-based and embedding-based SNA approaches. The results demonstrate that our proposed DPNIA outperforms nine baselines, whether attacking multiple networks simultaneously or just a single network. For example, when attacking two networks simultaneously, there is an average reduction of $13.6\%$ and a maximum reduction of $36.9\%$ compared to the best baselines under various conditions.
\end{itemize}

The organization of the subsequent sections in the paper is as follows: Section II provides an overview of related works in SNA and its attacks. Section III defines the problem related to node injection attacks. Section IV provides a detailed explanation of our proposed DPNIA. Section V presents our experimental results. Section VI concludes the paper and discusses potential future works.

\section{Related Works}
In this section, we provide an overview of the related works to SNA and the attacks on it.

\subsection{Methods of social network alignment}
Clues used to determine whether two accounts across different platforms belonging to the same user or not can be mined from user profile data, generated content data, social relationship data, etc. Thus, the main types of SNA methods can be divided into profile-based methods~\cite{MuXin-KDD2016,ZhaoDongsheng2018,li2019matching}, content-based methods~\cite{chen2020user,Riederer2016,feng2019dplink}, and network-structural-based methods~\cite{ZhouXiaoping2016,tang2020interlayer,ding2021soidp,ding2022user,LiuLi2016,ManTong2016-IJCAI,ZhouFan2018,liu2019structural,tang2023interlayer,zhang2023semi}. Among them, network-structural-based methods have attracted the attention of a growing number of scholars in the field in recent years. It can be categorized into neighborhood-based methods and embedding-based methods based on the main techniques employed for alignment~\cite{ShuKai2017}.

Neighborhood-based methods aim to capture the similarity score via the immediate neighbor nodes~\cite{ShuKai2017}. Narayanan and Shmatikov proposed the first method based on the node neighborhood to solve the SNA problem~\cite{Narayanan2009}. Later, Korula et al.~\cite{korula2014efficient} proposed an method to calculate the similarity score by counting the number of common matched neighbors. Zhou et al.~\cite{ZhouXiaoping2016} demonstrated the feasibility of using friend relationships for SNA through investigation. They also explored methods to address situations where a node shares the same number of common matched neighbors with other two nodes and conducted multiple rounds of iterations. Tang et al.~\cite{tang2020interlayer} introduced a degree penalty principle to distinguish the contribution of common matched neighbors for the SNA. Ding et al.~\cite{ding2021soidp} further explored the contribution played by second-order common matched neighbors. Meanwhile, they investigated the use of the naive Bayes model to measure the contributions in Ref.~\cite{ding2022user}. Zhu et al.~\cite{ZhuYuanyuan2012} modeled the SNA problem as a maximum common subgraph problem to obtain the aligning mapping. Zafarani and Liu~\cite{zafarani2015user} calculated the Laplacian matrices for each network and leveraged a matrix optimization method to perform the aligning. Zhang et al.~\cite{ZhangYutao2015} extended the alignment of two networks to propose a method for the comparison of global consistency among multiple social networks, enabling alignment between three or more social networks.

Embedding-based methods employ network embedding techniques to represent nodes in a dense space, ensuring that nodes with close network proximity exhibit similar vector representations~\cite{cai2018comprehensive}. Man et al.~\cite{ManTong2016-IJCAI} represented each social relationship network into a dense space and trained a cross-network mapping function for SNA. To improve the aligning accuracy, Zhou et al.~\cite{ZhouFan2018} proposed an method to pretrain the cross-network mapping function. Wang et al.~\cite{WangYongqing2019-www} addressed the efficiency issue of using embedding-based techniques for SNA. Different from represent each social network into an distinct space, Liu et al.~\cite{LiuLi2016} projected the two networks waiting for alignment into a shared space and then calculate the cosine similarity between the vectors of the unmatched nodes. They further improved the method by analyzing the influence of matched nodes from different communities in Ref.~\cite{liu2019structural}. Due to the expensive cost of labeling training data, Cheng et al.~\cite{cheng2019deep} proposed using active learning for SNA which only need to labeling the most crucial node pairs. Other works also cover aspects such as further enhance the accuracy~\cite{le2023enhancing}, multi-networks~\cite{ChuXiaokai2019-www}, multitasking~\cite{ren2020banana}, and heterogeneous network processing~\cite{wang2019user,zhou2019translink}.

\subsection{Attacks on social network alignment}
The key of prevent the leakage of personal privacy caused by malicious use of SNA is to reduce their aligning performance~\cite{shao2022adversarial}. One effective approach is NMA, which has been leveraged on the attacks to many network analysis scenarios, such as attacking node classification~\cite{dai2018adversarial,lin2023exploratory}, community detection~\cite{li2020adversarial,liu2022protect}, and hyperlink prediction~\cite{pan2021predicting,peng2022disintegrate,nie2023voluntary}.
Zhang et al.~\cite{zhang2020adversarial} proposed an adversarial attack model aimed at undermining the accuracy of SNA. This model pushes the nodes into dense regions, making it challenging to differentiate between the attacked nodes and those in densely areas. They adopted a similar strategy in Ref.~\cite{zhang2021adversarial}, and conducted a targeted analysis of the impact of gradient vanishing on training the attack model, proposing an attack signal amplification method to alleviate this effect. Tang et al.~\cite{TANG2022109095} proposed two strategies to searching the most important links for correct SNA and removed them hence degrade the aligning quality. Shao et al.~\cite{shao2023toak} designed a topological consistency breaking strategy to degrade SNA algorithms via removing links.

These methods provide insights into finding effective attacks. However, in social networks, where users own the nodes, the probability of getting a large number of users to add or remove friend connections according to the attackers' strategy is almost zero. Therefore, these methods are difficult to perform in practical scenarios. In the field of attacking the GNN, many researchers adopted the NIA, a more practical way. Wang et al.~\cite{wang2020scalable} proposed approximate fast gradient sign method to attack GNN in which adversaries can only inject new vicious nodes to the graphs. Sun et al.~\cite{sun2020adversarial} used the reinforcement learning to conduct NIA for the task of node classification task in a single graph. Hierarchical Q-learning network is adopted to manipulate the labels and the links between injected nodes and the chosen nodes. Zou et al.~\cite{zou2021tdgia} proposed an topological defective link selection strategy to choose the nodes that would be linked with the injected nodes. Meanwhile, they developed an smooth feature optimization method to generate the attribute features of the injected nodes. Tao et al.~\cite{tao2021single} focused on the extreme limitation that the attacker could only inject one single node at the test phase. Wang et al.~\cite{wang2022cluster} found that the NIA problem is equivalently formulated as a network clustering problem. Therefore, they proposed an method to solve the discrete optimization problem of the adjacency matrix of the attacked network in the context of network clustering.

Existing works of NIAs on the GNN tend to focus on attacking the tasks such as node classification and link prediction in a single network, and rarely consider scenarios involving multiple networks such as SNA attacks. Due to significant differences in the methods used to solve multiple network tasks and single network tasks, it is challenging to directly apply single network attack methods to multiple network tasks. In addition, from the perspective of representation learning, all nodes in a single network belong to a single latent space, while in multiple networks, nodes belong to multiple spaces, and each network's latent space is unknown to the others~\cite{ZhouFan2018}. Therefore, ASNA requires a targeted design and implementation.

\section{Preliminaries and problem}
In this section, we introduce the conception of the multiplex network for representing multiple OSN applications, and describe the problem of interlayer link prediction on the multiplex network. The symbols and notations frequently used in this paper are displayed in Table~\ref{tab:symbol}. We use bold uppercase letters for matrices, bold lowercase letters for vectors, and lowercase letters for scalars.

\subsection{Definitions}
In this section, we define the related terminologies and explain the problem of SNA and the node injection attack of SNA. The main symbols and notations used in this paper are shown in Table~\ref{tab:symbol}.
\begin{table}[!t]
\renewcommand{\arraystretch}{1.3}
\caption{Symbols and notations.}
\label{tab:symbol}
\centering
\begin{tabular}{|p{1cm}||p{6cm}|}
\hline
\textbf{Symbol }& \textbf{Description}\\
\hline
\hline
$G$ & A social network.\\
$v,V$ & The node and the node set, respectively.\\
$u$& An unmatched node.\\
$e,E$ & The intra-link and the intra-link set, respectively.\\
$\mathrm{A}$ & The adjacency matrix of $G$.\\
$\alpha, \mathrm{I}, \mathrm{II}$ & Social network indices.\\
$n$ & the number of nodes.\\
$\Delta$ & The budget.\\
$\Gamma$ & The neighbor set.\\
$\lambda$ & The vulnerability of a node.\\
$M$ &The matched neighbor set.\\
$C$ &The common matched neighbor set.\\
$U$ &The unmatched node set.\\
$\psi$&The correspondent node of a node or a correspondent node set of a node set.\\
$\Phi$&The set of the observed correspondent node pairs.\\
$\Psi$&The set of the unobserved correspondent node pairs.\\
$\Theta$&The set of the injected nodes.\\
$r$& A similarity score between two unmatched nodes across different networks.\\
\hline
\end{tabular}
\end{table}

\subsection{Social network alignment}
Given two social networks $G^I=(V^\mathrm{I},E^\mathrm{I})$ and $G^\mathrm{II}=(V^\mathrm{II},E^\mathrm{II})$ to be aligned, each of them is denoted as $G^\alpha=(V^\alpha,E^\alpha)(1\leq \alpha \leq 2)$. Notably, we use $G^\alpha$ to denote any social network. When given two social networks, we represent the specific first and second social networks as $G^I$ and $G^{II}$, respectively. In network $G^\alpha$, $V^\alpha={v^\alpha_1,\cdots,v^\alpha_i,\cdots, v^\alpha_{n^\alpha}}$ is the set of nodes, $E^\alpha=\{(v^\alpha_i,v^\alpha_j)\in V^\alpha\times V^\alpha |1\leq i,j \leq n^\alpha,i\neq j\}$ is the set of links. Since links in $E^\alpha$ are exist in $G^\alpha$, we call these links as \textbf{intra-links}. Additionally, $G^\alpha$ can be represented as adjacency matrix $A^\alpha$, where $A^\alpha(i,j)=1$ means that intra-link $v^\alpha_i,v^\alpha_j$ exists. If $A^\alpha(i,j)=A^\alpha(j,i)$ is always true, then $G^\alpha$ is an undirected network; otherwise, $G^\alpha$ is a directed network. In this paper, we focus on undirected graphs.

In some cases, two nodes come from different networks will belong to the same entity. For example, an account $v^\mathrm{I}_i$ in the social network Twitter belongs to the same user as an account $v^\mathrm{II}_j$ in the social network Facebook. To facilitate the analysis, usually, we can add a virtual link between such nodes to represent this relationship. Since such links are connected between $G^\mathrm{I}$ and $G^\mathrm{II}$, we can call them \textbf{inter-links}. Node $v^\mathrm{I}_i$ can be called as the \textbf{correspondent node} for $v^\mathrm{II}_j$, or vice versa. We can marked them as $\psi(v^\mathrm{I}_i)=v^\mathrm{II}_j$ or $\psi(v^\mathrm{II}_j)=v^\mathrm{I}_i$. The node pair ($v^\mathrm{I}_i$, $v^\mathrm{II}_j$) can be called \textbf{correspondent node pair}. If the correspondent node pair is known in advance, it is referred to as an \textbf{observed correspondent node pair}; otherwise, if it is not known in advance, it is an \textbf{unobserved correspondent node pair}. When leveraging SNA tools to align correspondent node pairs, if a correspondent node pair has been provided in advance or has been matched by the tools, we call it \textbf{matched node pair}. Nodes in matched node pairs are called matched nodes, while nodes that remain unaligned are called \textbf{unmatched nodes}.

\textbf{Definition 1: matched neighbor.} Given two nodes $v^\mathrm{I}_i$ and $v^\mathrm{II}_j$, if there is an intra-link between them and the correspondent node of $v^\alpha_j$ in $G^\mathrm{II}$ has been observed, we can call that the node $v^\alpha_j$ is the matched neighbor of $v^\alpha_i$.

\textbf{Definition 2: common matched neighbor.} Given a correspondent node pair ($v^\mathrm{I}_i$, $v^\mathrm{II}_j$), a node $v^\mathrm{I}_a$ in network $G^\mathrm{I}$ and a node $v^\mathrm{II}_b$ in network $G^\mathrm{II}$, if an intra-link exists between $v^\mathrm{I}_a$ and $v^\mathrm{I}_i$ and another intra-link exists between $v^\mathrm{II}_j$ and $v^\mathrm{II}_b$, we can say that the correspondent node pair ($v^\mathrm{I}_i$, $v^\mathrm{II}_j$) is the common matched neighbor of nodes $v^\mathrm{I}_a$ and $v^\mathrm{II}_j$. Furthermore, given a node set $V^\mathrm{I}_b$ in $G^\mathrm{I}$, if every node in it has an intra-link with $v^\mathrm{I}_a$, and $v^\mathrm{I}_a$ is a matched node, we can also refer to $v^\mathrm{I}_a$ as the common matched neighbor of all nodes in $V^\mathrm{I}_b$.

Formally, Given two social networks $G^\mathrm{I}$ and $G^\mathrm{II}$, the node sets $V^\mathrm{I}$ and $V^\mathrm{II}$, intra-links sets $E^\mathrm{I}$ and $E^\mathrm{II}$, the observed inter-links set $\Phi$, a SNA method need to learns a binary function $f$: $U^\mathrm{I} \times U^\mathrm{II} \rightarrow 0,1$ such that
\begin{equation}
f(u^\mathrm{I}_i,u^\mathrm{II}_j) = \left\{
\begin{array}{ll}
1, & \mbox{if an inter-link exists} \\
0, & \mbox{otherwise}
\end{array},
\right.
\label{eq:objectivefunction}
\end{equation}
where $U^\mathrm{I}$ and $U^\mathrm{II}$ are the unmatched node sets of $G^\mathrm{I}$ and $G^\mathrm{II}$, respectively.

\subsection{Node injection attack}
Several studies have investigated the network modification attack which generates perturbation by modifying the network structure $A^\alpha$ of the original social network $G^\alpha$. The modification methods can be categorized into three types including adding, deleting, and modifying intra-links within the set $E^\alpha$. In this paper, we explore a novel approach to attack the SNA tools or models which can be called node injection attack (NIA). This approach injects new nodes into $G^\alpha$ while keeping the original intra-links in $E^\alpha$ unchanged to generate perturbation. Formally, NIA constructs ${G^\alpha}'$ within a budget $\Delta^\alpha$ by
\begin{equation}
{A^\alpha}' = \left[
\begin{array}{ll}
{A^\alpha} & {A^\alpha}_{atk}^{T} \\
{A^\alpha}_{atk} & {I^\alpha}_{atk}
\end{array}
\right],
\label{eq:objectivefunction}
\end{equation}
where $A^\alpha_{atk} \in \{0,1\}$ denotes the intra-links between the injected nodes and the existing nodes while $I^\alpha_{atk} \in \{0,1\}$ denotes the intra-links within the injected nodes.
The attacks can be performed on $A^\alpha_{atk}$ and $I^\alpha_{atk}$. Since that the cost of adding intra-links between two injected nodes is much smaller than adding intra-links between an injected node and an existing node,
the constraint of node injection attack can be expressed as
\begin{equation}
|{A^\alpha}_{atk}|\leq \Delta^\alpha \in Z.
\end{equation}
It center on two primary aspects: the number of injected nodes and the number of intra-links added by the injected nodes.

The purpose of the constraint is to ensure the usability of the social network platform. Unrestricted injection, although effective, may potentially overwhelm authentic information with false data. Simultaneously, there are considerations related to cost and avoiding detection by the defender.

The objective function of the attack is to minimize the number of correct alignment by injecting nodes within the budget:
\begin{equation}
\begin{array}{c}
\min\limits_{{A^\mathrm{I}}',{A^\mathrm{II}}'}(\sum\limits_{(u^\mathrm{I}_i,u^\mathrm{II}_j) \in \Psi}{\mathds{1}(f(u^\mathrm{I}_i,u^\mathrm{II}_j)=1)})\\
s.t.~|A^\mathrm{I}_{atk}|\leq \Delta^\mathrm{I},|A^\mathrm{II}_{atk}|\leq \Delta^\mathrm{II}
\end{array},
\label{eq_objfun_atk}
\end{equation}
where $\Psi$ is the set of unobserved correspondent node pairs.

\section{Methodology}
\begin{figure*} [t!]
    \centering
    \includegraphics[width=0.99\textwidth]{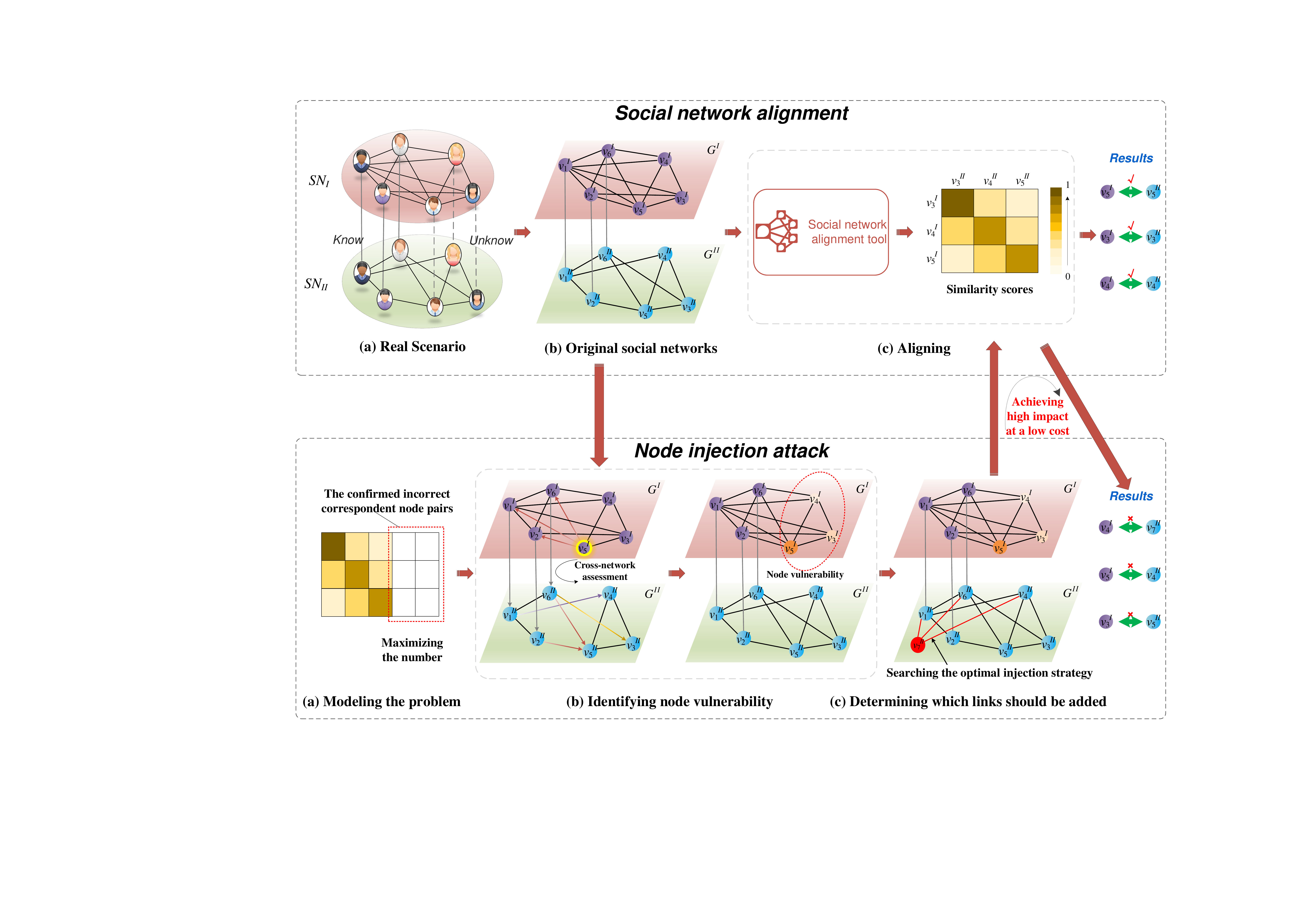}
    \caption{An overview of the proposed low-cost, high-impact node injection attack framework for ASNA.}
    \label{pic_nodeinfectionattack}
\end{figure*}

Figure~\ref{pic_nodeinfectionattack} illustrates our proposed DPNIA framework. It models ASNA as a problem of maximizing the number of confirmed incorrect correspondent node pairs who have a greater similarity scores than the pairs between existing nodes, making ASNA solvable. Then, it employs the designed cross-network evaluation method to identify node vulnerability, facilitating a progression of attacks from easy to difficult. Finally, it utilizes the developed optimal injection strategy searching method, based on dynamic programming, to determine which links should be added between injected nodes and existing nodes, thereby achieving a high impact for attack effectiveness at a low cost.

\subsection{Modeling the problem}
Given the social networks $G^\mathrm{I}$ and $G^\mathrm{II}$, when dealing with an unmatched node $u^\mathrm{I}_i$, SNA algorithms or models often calculate similarity scores $\{r(u^\mathrm{I}_i,u^\mathrm{II}_k)|u^\mathrm{II}_k \in U^\mathrm{II}\}$ between $u^\mathrm{I}_i$ and each unmatched node in $G^\mathrm{II}$. These similarity scores are then ranked, and the node with the highest score is chosen as the correspondent node for $u^\mathrm{I}_i$.

To carry out an attack, we can make the similarity score between $u^\mathrm{I}_i$ and its true correspondent node $u^\mathrm{II}_j$ less than one or more of the similarity scores between $u^\mathrm{I}_i$ and other nodes in $G^\mathrm{II}$. Assuming $(\psi(u^\mathrm{I}_i)=u^\mathrm{II}_j)$, the objective function can transition from Eq. (\ref{eq_objfun_atk}) to
\begin{equation}
\begin{array}{c}
\max\limits_{{A^\mathrm{II}}'}(\sum\limits_{(u^\mathrm{I}_i,\psi(u^\mathrm{I}_i)) \in \Psi} \sum\limits_{\substack{u^\mathrm{II}_k \in U^\mathrm{II},\psi(u^\mathrm{I}_i)\neq u^\mathrm{II}_k}} {\mathds{1}(r(u^\mathrm{I}_i,\psi(u^\mathrm{I}_i))<r(u^\mathrm{I}_i,u^\mathrm{II}_k))})\\
s.t.~|A^\mathrm{II}_{atk}|\leq \Delta^\mathrm{II}
\end{array}.
\label{eq_objfun_atk2}
\end{equation}
The objective function for injected nodes in $G^\mathrm{I}$ is similar.

However, since $\Psi$ is unknown, we are uncertain about which node in $G^\mathrm{II}$ is real $\psi(u^\mathrm{I}_i)$ for $u^\mathrm{I}_i$. An alternative approach is to make the injected node have a similarity score with $u^\mathrm{I}_i$ greater than the maximum similarity score between $u^\mathrm{I}_i$ and all unmatched nodes in $G^\mathrm{II}$. In other words, the goal can be transition to
\begin{equation}
\begin{array}{c}
\max\limits_{{A^\mathrm{II}}'}(\sum\limits_{u^\mathrm{I}_i \in U^\mathrm{I}}\sum\limits_{u^\mathrm{II}_a \in \Theta^\mathrm{II}} {\mathds{1}(\max(\{r(u^\mathrm{I}_i,u^\mathrm{II}_k)\})<r(u^\mathrm{I}_i,u^\mathrm{II}_a))})\\
s.t.~|A^\mathrm{II}_{atk}|\leq \Delta^\mathrm{II}, u^\mathrm{II}_k \in U^\mathrm{II}
\end{array},
\label{eq_objfun_atk_inject}
\end{equation}
where $\Theta^\mathrm{II}$ is the set of injected nodes for $G^\mathrm{II}$.

By the above way, ASNA is modeled as a problem of maximizing the number of confirmed incorrect correspondent node pairs who have a greater similarity scores than the pairs between existing nodes. This method effectively disrupts the discrimination process of SNA algorithms or models, making the problem solvable.

\subsection{Identifying node vulnerability}
Now, we need to answer how to achieve the goal modeled by Eq.~(\ref{eq_objfun_atk_inject}). Intuitively, if the budget is limited, we could prioritize attacking nodes in the network that are more susceptible to be perturbed. We use the concept of vulnerability to describe the degree to which a node is susceptible. Nodes with higher vulnerability are more susceptible to alignment attacks, necessitating lower costs; while nodes with lower vulnerability are more resistant to alignment attacks, necessitating higher costs. Clearly, we need a method to distinguish the vulnerability of different nodes and then find the optimal attack strategy.

Next, we conduct an analysis of node vulnerability in SNA to discover ways to differentiate these vulnerabilities. As we known, various methods have been proposed for calculating similarity scores, such as computing a similarity score between nodes using a proposed method~\cite{ZhouXiaoping2016,tang2020interlayer,ding2021soidp,ding2022user} or measuring distance or cosine similarity in the embedding space~\cite{LiuLi2016,ManTong2016-IJCAI,ZhouFan2018,liu2019structural,tang2023interlayer,zhang2023semi}.

For neighborhood-based SNA methods, the analysis of their similarity score calculation reveals that $r(u^\mathrm{I}_i,u^\mathrm{II}_k)$ is typically proportional to the number of common matched neighbors between $u^\mathrm{I}_i$ and $u^\mathrm{II}_k$. For instance, the similarity score in Ref.~\cite{ding2022user} is
\begin{equation}
r(u^\mathrm{I}_i,u^\mathrm{II}_k)=\sum_{\sigma \in C{(u^\mathrm{I}_i,u^\mathrm{II}_k)}} {ln(sR_\sigma^*)},
\end{equation}
where $C{(u^\mathrm{I}_i,u^\mathrm{II}_k)}$ represents the common matched neighbor set for node pair $(u^\mathrm{I}_i$, $u^\mathrm{II}_k)$. The meanings of $s$ and $R_\sigma^*$ can be found in Ref.~\cite{ding2022user}. The similarity score in Ref.~\cite{ZhouXiaoping2016} is
\begin{equation}
r(u^\mathrm{I}_i,u^\mathrm{II}_k)=|C(u^\mathrm{I}_i,u^\mathrm{II}_k)|+\frac{|C(u^\mathrm{I}_i,u^\mathrm{II}_k)|}{min(|\Gamma(u^\mathrm{I}_i)|,|\Gamma(u^\mathrm{II}_k)|)},
\end{equation}
where $\Gamma(\cdot)$ represents the set of neighbor for a node. The similarity score in Ref.~\cite{tang2020interlayer} is
\begin{equation}
\begin{array}{l}
r(u^\mathrm{I}_i,u^\mathrm{II}_k)=\\
\sum\limits_{(u^\mathrm{I}_a,u^\mathrm{II}_b) \in C{(u^\mathrm{I}_i,u^\mathrm{II}_k)}} {\log^{-1}(|\Gamma(u^\mathrm{I}_a)|+1) \cdot \log^{-1}(|\Gamma(u^\mathrm{II}_b)|+1) }.
\end{array}
\end{equation}

For these methods, let's examine the scenario where $u^\mathrm{I}_i$ and $u^\mathrm{II}_k$ have a common matched neighbor count of $a$, while $u^\mathrm{I}_i$ and $u^\mathrm{II}_j$ have a larger common matched neighbor count of $b_1$ ($a < b_1$). To make $r(u^\mathrm{I}_i,u^\mathrm{II}_k) > r(u^\mathrm{I}_i,u^\mathrm{II}_j)$, a perturbation budget $\Delta_1$ is required such that $a + \Delta_1 = b_1 + \epsilon$, where $\epsilon$ represents the minimum cost needed to achieve an increase in the similarity score to match $r(u^\mathrm{I}_i,u^\mathrm{II}_j)$.

If the number of common matched neighbors between $u^\mathrm{I}_i$ and $u^\mathrm{II}_j$ is greater than $b_1$ and denoted as $b_2$, then to ensure that $a+\Delta_2=b_2+\epsilon$, the required perturbation budget $\Delta_2$ must indeed be greater than $\Delta_1$. Therefore, the more common matched neighbors $u^\mathrm{II}_j$ shares with $u^\mathrm{I}_i$, the higher the cost required to make $r(u^\mathrm{I}_i,u^\mathrm{II}_k) > r(u^\mathrm{I}_i,u^\mathrm{II}_j)$. Consequently, the vulnerability of $u^\mathrm{I}_i$ is lower when it has more common matched neighbors with $u^\mathrm{II}_j$, and conversely, its vulnerability is higher when it has fewer common matched neighbors with $u^\mathrm{II}_j$.

\begin{figure} [t!]
    \centering
    \includegraphics[width=0.49\textwidth]{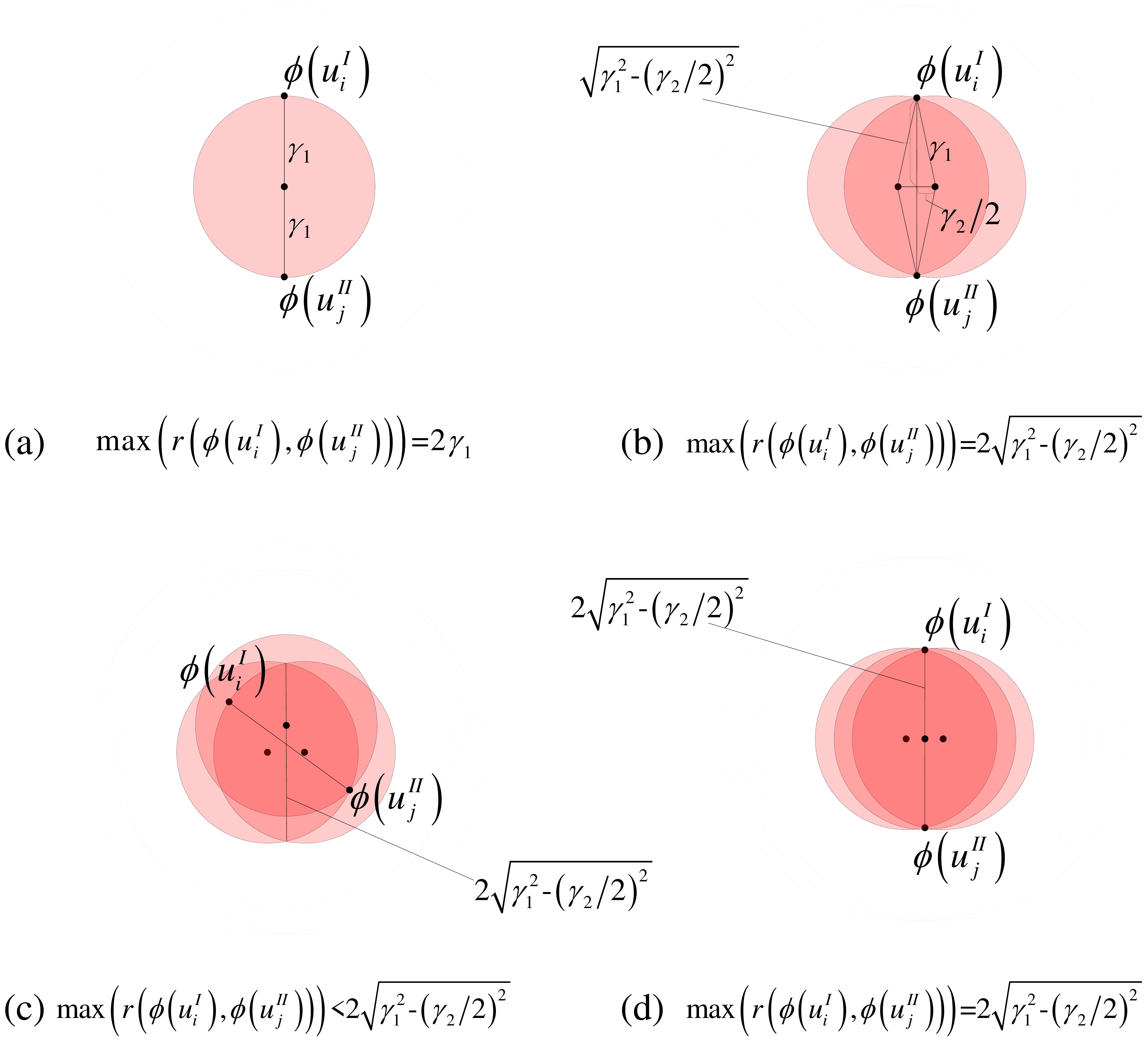}
    \caption{Example of the maximum distance between two unmatched nodes in the embedding space with different number of common matched neighbors.}
    \label{pic_vulnerability}
\end{figure}

For embedding-based SNA methods, we can unify the methods used in related literature for calculating the similarity scores of the embedding vectors into a common format $r(u^\mathrm{I}_i,u^\mathrm{II}_j)=1/||\phi(u^\mathrm{I}_i),\phi(u^\mathrm{I}_j)||_2$, where $||\cdot||_2$ represents the L2 norm, and $\phi(\cdot)$ represents the mapping of nodes into an embedding space. The Ref.~\cite{cai2018comprehensive} has explained that network embedding techniques aim to map nodes into an embedding space in a way that similar nodes in the embedding space are closer together. Suppose two nodes in a social network with intra-links have a maximum distance of $\gamma_1$ in the embedding space, and the lower bound of the distance between two nodes in the embedding space is $\gamma_2$. Obviously, the maximum distance between two matched neighbors of a node is $2\gamma_1$. As shown in Figure~\ref{pic_vulnerability} (a), if $u^\mathrm{I}_i$ and $u^\mathrm{II}_j$ share one common matched neighbor, the maximum value of $||\phi(u^\mathrm{I}_i)-\phi(u^\mathrm{II}_j)||_2$ is $2\gamma_1$. This is because SNA models based on network embeddings aim to make nodes in the training set overlap as much as possible in the embedding space, meaning common matched neighbors of $u^\mathrm{I}_i$ and $u^\mathrm{II}_j$ will overlap in the embedding space.

As shown in Figure~\ref{pic_vulnerability} (b), if $u^\mathrm{I}_i$ and $u^\mathrm{II}_j$ have two common matched neighbors, the maximum value of $||\phi(u^\mathrm{I}_i)-\phi(u^\mathrm{II}_j)||_2$ will be $2\sqrt{\gamma_1^2-(\gamma_2/2)^2}$. If $\gamma_2>0$, then $2\sqrt{\gamma_1^2-(\gamma_2/2)^2}>2\gamma_1$. As shown in Figure~\ref{pic_vulnerability} (c) and (d), if $u^\mathrm{I}_i$ and $u^\mathrm{II}_j$ have three common matching neighbors, the maximum value of $||\phi(u^\mathrm{I}_i)-\phi(u^\mathrm{II}_j)||_2$ will either decrease or remain the same. This implies that as the number of common matched neighbors between $u^\mathrm{I}_i$ and $u^\mathrm{II}_j$ increases, the maximum value of $||\phi(u^\mathrm{I}_i)-\phi(u^\mathrm{II}_j)||_2$ can only decrease or remain the same, but it cannot increase. Consequently, $1/||\phi(u^\mathrm{I}_i)-\phi(u^\mathrm{II}_j)||_2$ is more likely to have a larger value as the number of common matched neighbors increases.

In other words, as the number of common matched neighbors increases, the similarity score is more likely to have a larger value. Assuming $r(u^\mathrm{I}_i,u^\mathrm{II}_j)$ is greater than $r(u^\mathrm{I}_i,u^\mathrm{II}_k)$, if the number of common matched neighbors between $u^\mathrm{I}_i$ and $u^\mathrm{II}_j$ increases, $r(u^\mathrm{I}_i,u^\mathrm{II}_j)$ is more likely to have a larger value. Therefore, increasing $r(u^\mathrm{I}_i,u^\mathrm{II}_k)$ above $r(u^\mathrm{I}_i,u^\mathrm{II}_j)$ through perturbation may require a higher cost. Hence, for methods based on network embeddings, the more common matched neighbors $u^\mathrm{II}_j$ shares with $u^\mathrm{I}_i$, the higher the potential cost may be to make $r(u^\mathrm{I}_i,u^\mathrm{II}_k) > r(u^\mathrm{I}_i,u^\mathrm{II}_j)$, which may result in lower vulnerability for $u^\mathrm{I}_i$. Conversely, higher vulnerability may occur when fewer common matched neighbors are present.

Additionally, the difference between the number of matched neighbors and the number of common matched neighbors also impacts the vulnerability of existing nodes. For the neighborhood-based methods, if the number of matched neighbors is equal to the number of common matched neighbors, it implies that it is difficult to make $r(u^\mathrm{I}_i,u^\mathrm{II}_k)$ greater than $r(u^\mathrm{I}_i,u^\mathrm{II}_j)$. For embedding-based methods, it means there are no additional reference points available to enable $u^\mathrm{II}_k$ to get closer to $u^\mathrm{I}_i$ than $u^\mathrm{II}_j$.

Based on the analysis above, using $M(u_i^\mathrm{I})$ to represent the set of matched neighbors of $u_i^\mathrm{I}$, we can calculate the vulnerability of a node by
\begin{equation}
\begin{array}{l}
\lambda_i^\mathrm{I}=\delta \cdot \frac{|M(u_i^\mathrm{I})|-max(\{|C(u_i^\mathrm{I},u_k^\mathrm{II})||u_k^\mathrm{II} \in U^\mathrm{II}\})}{|M(u_i^\mathrm{I})|+1} \\
+ (1-\delta) \cdot \frac{1}{max(\{|C(u_i^\mathrm{I},u_k^\mathrm{II})||u_k^\mathrm{II} \in U^\mathrm{II}\})+1}.
\end{array}
\label{eq_lambda}
\end{equation}
In this equation, $\delta$ is a hyperparameter used to distinguish the importance of different influencing factors. The formula ensures that among the neighbors, the more matched neighbors a node has, the more susceptible it is to precise attacks, and thus, the more vulnerable it is. Conversely, the more common matched neighbors a node has, the higher the cost required for a precise attack, making it less vulnerable. Additionally, when there is a larger difference between the number of matched neighbors and the number of common matched neighbors, there is a greater opportunity to make the similarity scores of other node pairings exceed the correct pairing's similarity score. In this way, we only need to ensure that the similarity scores of confirmed incorrect pairs are greater than those of any other pairs through node injection. There is no need for repeated comparisons of the impact of ASNA on all node pairs across different social networks after injection, enabling the problem to be solved within a limited time. When measuring the vulnerability of each unmatched node, the method we propose takes into account various factors such as unmatched nodes, their matched neighbors, possible counterpart nodes in the other network, possible counterpart nodes' matched neighbors, and their common matched neighbors. Accurate sensing and calculations across both networks are required. Therefore, we refer to this approach as cross-network assessment.

\subsection{Determining which links should be added}
After obtaining the vulnerability of different unmatched nodes, the next step is how to inject nodes to the network. We develop an optimal injection strategy searching method based on dynamic programming to determine which intra-links should be added between injected nodes and existing nodes, thereby achieving a high impact for attack effectiveness at a low cost.

First, we sort the vulnerability of all unmatched nodes in $G^\mathrm{I}$ in descending order, prioritizing the most vulnerable nodes. Then, we sequentially taking attacks on the unmatched nodes, ensuring that these nodes achieve the maximum similarity scores with non-corresponding nodes in $G^\mathrm{II}$. The budget allocated to attacking $u^\mathrm{I}_i$ in $G^\mathrm{II}$ is determined as follows:
\begin{equation}
\begin{array}{l}
\Delta^\mathrm{II}_{u^\mathrm{I}_i}=\max(\{|C(u_i^\mathrm{I},u_k^\mathrm{II})||u_k^\mathrm{II} \in U^\mathrm{II}\})\\
+\mathrm{sign}(|M(u_i^\mathrm{I})|-\max(\{|C(u_i^\mathrm{I},u_k^\mathrm{II})||u_k^\mathrm{II} \in U^\mathrm{II}\})),
\end{array}
\label{eq_delta_uIi}
\end{equation}
where $\mathrm{sign}(\cdot)$ represents the sign function. With Eq.~(\ref{eq_delta_uIi}), injected nodes only need to link to $\Delta^\mathrm{II}_{u^\mathrm{I}_i}$ correspondent nodes of the matched neighbors for $u^\mathrm{I}_i$. We employ this way to ensure that the injected node achieves a higher similarity score with $u^\mathrm{I}_i$ than $r(u^\mathrm{I}_i,\psi(u^\mathrm{I}_i))$, while minimizing the budget required. When the number of common matched neighbors between $u^\mathrm{I}_i$ and $\psi(u^\mathrm{I}_i)$ is $\max({|C(u_i^\mathrm{I},u_k^\mathrm{II})||u_k^\mathrm{II} \in U^\mathrm{II}})$, if an injected node links to $\Delta{u^\mathrm{I}_i}-1$ correspondent nodes of $u^\mathrm{I}_i$'s matched neighbors, the similarity score between the injected node and $u^\mathrm{I}_i$ can, at best, be equal to $r(u^\mathrm{I}_i,\psi(u^\mathrm{I}_i))$. Therefore, Eq.~(\ref{eq_delta_uIi}) ensures that injected nodes can successfully perturb node $u^\mathrm{I}_i$ with the minimum cost.

While the above method allows injected nodes to successfully disrupt single node's alignment at the lowest possible cost, the budget remains limited for the task of aligning the entire social networks. Is it possible to make the addition of $\Delta^\mathrm{II}_{u^\mathrm{I}_i}$ intra-links disrupt the alignment of more unmatched nodes? Can a method be designed to find the maximum number of unmatched nodes that can be disrupted and the corresponding strategy to achieve this maximum value?

To address the issue mentioned above, we can solve it in the following manner. Let $U^\mathrm{I}_{xy}=\{u^\mathrm{I}_i|i \in [1,x]\}$ represents a set composed of $x$ unmatched nodes, and $C^\mathrm{I}_{xy}$ represents the set of common matched neighbors of all nodes in $U^\mathrm{I}_{xy}$ where $y$ is an index to distinguish different sets. If, after adding intra-links from injected node $u^\mathrm{II}_a$ for all nodes in $\psi(C^\mathrm{I}_{xy})$, for any $u^\mathrm{I}_{i}\in U^\mathrm{I}_{xy}$, we have $r(u^\mathrm{I}_{i},u^\mathrm{II}_a)>\max(\{r(u^\mathrm{I}_{i},u^\mathrm{II}_k)|u^\mathrm{II}_k\in U^\mathrm{II}\})$, then we can achieve a successful attack on all nodes in $U^\mathrm{I}_{xy}$ with a budget of $\max(\{\Delta^\mathrm{II}_{u^\mathrm{I}_{i}}\})$, without the need for a budget as large as $\sum_{u^\mathrm{I}_{i}\in U^\mathrm{I}_{xy}}{\Delta^\mathrm{II}_{u^\mathrm{I}_{i}}}$.

To achieve maximum disruption at the minimum cost, we need to find the set of $\psi(C^\mathrm{I}_{xy})$ that maximizes the value of $x$. Additionally, if $\Delta^\mathrm{II}$ is significantly larger than $\max(\{\Delta^\mathrm{II}_{u^\mathrm{II}_{i}}\})$, we can continue to search for the second-largest, third-largest, and more sets until $\Delta^\mathrm{II}$ is exhausted.
The solution to this problem satisfies the properties of optimal substructure and overlapping subproblems. The search for the correspondent nodes of the common matched neighbors of all unmatched nodes in the set $U^\mathrm{I}_{xy}$ can be decomposed into several subsets' search for the common matched neighbors of unmatched nodes and then taking their intersection. Additionally, within any given subset, the correspondent nodes of all unmatched nodes' common matched neighbors can also be utilized for the solution of another set, denoted as $U^\mathrm{I}_{xz}$, to which the subset simultaneously belong.

For node $u^\mathrm{I}_i$ in $G^\mathrm{I}$, we compute the minimum budget that can be allocated to ensure the success of the attack using Eq.~(\ref{eq_delta_uIi}). In the equation, the calculation formula for the number of common matching neighbors between $u^\mathrm{I}_i$ and $u^\mathrm{II}_k$ is given by
\begin{equation}
|C(u^\mathrm{I}_i,u^\mathrm{II}_k)|=\sum\limits_{\substack{(v^\mathrm{I}_a,v^\mathrm{II}_a)\in \Phi ,\\v^\mathrm{I}_a \in \Gamma(u^\mathrm{I}_i),\\ v^\mathrm{II}_a \in \Gamma(u^\mathrm{II}_k)}}{e^\mathrm{I}_{ia}\cdot e^\mathrm{II}_{aj}},
\label{eq_cal_cmn_corssnetworks}
\end{equation}
where $e^\mathrm{I}_{ia}$ represents whether there exists an intra-link between $u^\mathrm{I}_i$ and $v^\mathrm{I}_a$. $e^\mathrm{I}_{ia}=1$ indicates that there exists an intra-link between $u^\mathrm{I}_i$ and $v^\mathrm{I}_a$, while $e^\mathrm{I}_{ia}=0$ indicates that there does not exist an intra-link between $u^\mathrm{I}_i$ and $v^\mathrm{I}_a$. The reason for using the same subscript "a" for $v^\mathrm{I}_a$ and $v^\mathrm{II}_a$ is that $(v^\mathrm{I}_a,v^\mathrm{II}_a)$ is an correspondent node pair and their relationship is one-to-one.

Eq.~(\ref{eq_cal_cmn_corssnetworks}) can be expressed in the form of a vector product as follows:
\begin{equation}
\begin{array}{l}
|C(u^\mathrm{I}_i,u^\mathrm{II}_k)|=[e^\mathrm{I}_{i1},\cdots,e^\mathrm{I}_{ia},\cdots,e^\mathrm{I}_{i|\Phi|}] \cdot
\left[
 \begin{array}{c}
 e^\mathrm{II}_{1k} \\
 \vdots\\
 e^\mathrm{II}_{ak} \\
 \vdots\\
 e^\mathrm{II}_{|\Phi|k} \\
 \end{array}
\right]\\
\end{array}.
\label{eq_cal_cmn_corssnetworks_vec}
\end{equation}
Let $\mathbf{e}^\mathrm{I}_i=[e^\mathrm{I}_{i1},\cdots,e^\mathrm{I}_{ia},\cdots,e^\mathrm{I}_{i|\Phi|}]^T$ and $\mathbf{e}^\mathrm{II}_k=[e^\mathrm{II}_{1k},\cdots,e^\mathrm{II}_{ak},\cdots,e^\mathrm{II}_{|\Phi|k}]^T$, Eq.~(\ref{eq_cal_cmn_corssnetworks_vec}) can be simplified as
\begin{equation}
|C(u^\mathrm{I}_i,u^\mathrm{II}_k)|=(\mathbf{e}^\mathrm{I}_i)^T \cdot \mathbf{e}^\mathrm{II}_k,
\label{eq_cal_cmn_corssnetworks_simple}
\end{equation}
where $(\cdot)^T$ represents the transpose of a vector or matrix.

Supposing $\mathbf{E}^\mathrm{I}=[(\mathbf{e}^\mathrm{I}_1)^T,\cdots,(\mathbf{e}^\mathrm{I}_i)^T,\cdots,(\mathbf{e}^\mathrm{I}_{|U^\mathrm{I}|})^T]$ and $\mathbf{E}^\mathrm{II}=[(\mathbf{e}^\mathrm{II}_1)^T,\cdots,(\mathbf{e}^\mathrm{II}_k)^T,\cdots,(\mathbf{e}^\mathrm{II}_{|U^\mathrm{II}|})^T]$, the numbers of common matched neighbors between all unmatched nodes in $G^\mathrm{I}$ and all unmatched nodes in $G^\mathrm{II}$ can be represented as
\begin{equation}
\mathbf{C}=\mathbf{E}^\mathrm{I} \cdot (\mathbf{E}^\mathrm{II})^T.
\label{eq_cal_cmn_crossnetworks_mat}
\end{equation}
The element in the $i$th row and $j$th column of matrix $\mathbf{C}$,  denoted as $\mathbf{C}(i, k)$, is the value of $|C(u^\mathrm{I}_i,u^\mathrm{II}_k)|$.

In vector $\mathbf{e}^\mathrm{I}_i$, if the element in the $a$th row is 1, it means that $v^\mathrm{I}_a$ is a matched neighbor of $u^\mathrm{I}_i$. Therefore, we can directly use vector $\mathbf{e}^\mathrm{I}_i$ to quickly find the set $M(u^\mathrm{I}_i)$. Supposing $\mathbf{l}=[1,1,\cdots,1]_{1\times|\Phi|}^T$, then
\begin{equation}
|M(u^\mathrm{I}_i)|=(\mathbf{e}^\mathrm{I}_i)^T \cdot \mathbf{l}.
\end{equation}
Using $M^\mathrm{I}(:)$ to represent the vector composed of the number of matched neighbors for each unmatched node in $G^\mathrm{I}$, we have:
\begin{equation}
|M^\mathrm{I}(:)|=\mathbf{E}^\mathrm{I} \cdot \mathbf{l}.
\label{eq_cmn_intranetworks_num}
\end{equation}

With the help of Eq.~(\ref{eq_cal_cmn_crossnetworks_mat}) and Eq.~(\ref{eq_cmn_intranetworks_num}), the budget for each unmatched node in $G^\mathrm{I}$ can be easily computed. For any node $u^\mathrm{I}_i$, we need to select $\Delta^\mathrm{II}_{u^\mathrm{I}_i}$ nodes from the set $\psi(M(u^\mathrm{I}_i))$. These selected nodes will be added intra-links with an injected node. To maximize the impact of the budget of $\Delta^\mathrm{II}_{u^\mathrm{I}_i}$ on more nodes, the key is to identify nodes that have a common matched neighbor count with $u^\mathrm{I}_i$ greater than or equal to $\Delta^\mathrm{II}_{u^\mathrm{I}_i}$ and then instruct the injected nodes to add the correspondent nodes of these common matched neighbors.

Suppose the nodes in $U^\mathrm{I}_{xy}$ are $\{u^\mathrm{I}_1,u^\mathrm{I}_2\}$. Whether the matched node $v^\mathrm{I}_a$ is a common matched neighbor of these two nodes can be determined by $e^\mathrm{I}_{1a} \cdot e^\mathrm{I}_{2a}$. Therefore, for all matched nodes in $G^\mathrm{I}$, whether they are common matched neighbors of $u^\mathrm{I}_1$ and $u^\mathrm{I}_2$ can be determined using $\mathbf{e}^\mathrm{I}_1 \circ \mathbf{e}^\mathrm{I}_2$. If the $i$-th element of the resulting vector has a value of 1, it means that $v^\mathrm{I}_i$ is a common matched neighbor of $u^\mathrm{I}_1$ and $u^\mathrm{I}_2$. The count of common matched neighbors between $u^\mathrm{I}_1$ and $u^\mathrm{I}_2$ can be calculated as $(\mathbf{e}^\mathrm{I}_1)^T \cdot \mathbf{e}^\mathrm{I}_2$. When the nodes in $U^\mathrm{I}{xy}$ are ${u^\mathrm{I}_1,u^\mathrm{I}_2,u^\mathrm{I}_3}$, for all matched nodes in $G^\mathrm{I}$, whether they are common matched neighbors of these three nodes can be calculated using $\mathbf{e}^\mathrm{I}_1 \circ \mathbf{e}^\mathrm{I}_2\circ \mathbf{e}^\mathrm{I}_3$. Meanwhile, the formula for calculating the count of common matched neighbors of these three nodes is $((\mathbf{e}^\mathrm{I}_1 \circ \mathbf{e}^\mathrm{I}_2)^T \cdot \mathbf{e}^\mathrm{I}_3$.

By extension, when the nodes in $U^\mathrm{I}_{xy}$ are ${u^\mathrm{I}_1,u^\mathrm{I}_2,\cdots, u^\mathrm{I}_x}$, the common matched neighbors of these nodes can be found by

\begin{equation}
\begin{array}{c}
CMN(U^\mathrm{I}_{xy},:)=\mathbf{e}^\mathrm{I}_1 \circ~\cdots~\circ \mathbf{e}^\mathrm{I}_{x-1} \circ \mathbf{e}^\mathrm{I}_{x}\\
=CMN(U^\mathrm{I}_{x-1y'},:)\circ \mathbf{e}^\mathrm{I}_{x}
\end{array}.
\label{eq_cmn_intranetworks}
\end{equation}
The equation for calculating their common matched neighbors is
\begin{equation}
\begin{array}{c}
NC(U^\mathrm{I}_{xy})=(\mathbf{e}^\mathrm{I}_1 \circ~\cdots~\circ \mathbf{e}^\mathrm{I}_{x-1})^T \cdot \mathbf{e}^\mathrm{I}_{x}\\
=CMN(U^\mathrm{I}_{x-1y'},:)^T\cdot \mathbf{e}^\mathrm{I}_{x}
\end{array},
\label{eq_number_cmn}
\end{equation}
where $y'$ represents an index value different with $y$.

In the actual solving process, we can start with any set of two unmatched nodes, calculate their common matched neighbors according to Eq.~(\ref{eq_number_cmn}), and then compare it with the minimum budget obtained from Eq.~(\ref{eq_delta_uIi}), Eq.~(\ref{eq_cal_cmn_crossnetworks_mat}), and Eq.~(\ref{eq_cmn_intranetworks_num}). If it exceeds this budget, the search continues using Eq.~(\ref{eq_cmn_intranetworks}) and Eq.~(\ref{eq_number_cmn}), until it is less than this budget. During the search, it's necessary to keep a record of already computed node sets to avoid extensive duplicate calculations, making the problem tractable.

By taking the above strategy, we can optimize the injection attacks to maximize the perturbation to SNA while working within a limited budget.

\begin{table}[!th]
\centering
\caption{Basic statistical information of the datasets including the number of nodes $|V|$, the number of intra-links $|E|$, and the number of inter-links $|\Phi|$.}
\label{table}
\setlength{\tabcolsep}{3pt}
\begin{tabular}{cccc}
\hline
Network&$|V|$&$|E|$&$|\Phi|$\\
\hline
Foursquare1&5,313&76,972&\multirow{2}*{3,148}\\
Twitter1&5,120&164,920&~\\
\hline
Foursquare2&1,507&18,470&\multirow{2}*{1,507}\\
Twitter2&1,507&13,843&~\\
\hline
Higgs\_Friendship&4,288&122,826&\multirow{2}*{3,760}\\
Higgs\_Mention&3,777&13,413&~\\
\hline
Higgs\_Friendship&4,184&101,618&\multirow{2}*{3,219}\\
Higgs\_Retweet&3,238&13,571&~\\
\hline
\end{tabular}
\label{tab:datasets}
\end{table}
\section{Experiments}
In this section, we first introduce the experimental configurations including datasets, baselines, SNA approaches, evaluation metrics, and other details of the experiments. Then, we compare the proposed attacking method with nine baselines by several experiments on different settings. To effectively evaluate the performance of these network alignment attacking methods, each of them is deployed to attack five well known network alignment approaches on several public real-world datasets.

\subsection{Experimental configurations}
\textbf{Datasets.} We conduct our experiments on four public real-world datasets. They are Foursquare-Twitter1 (\textbf{FT1})~\cite{ZhangJiawei2015-IJCAI},  Foursquare-Twitter2 (\textbf{FT2})~\cite{jalili2017link}, Higgs\_Friendship-Mention (\textbf{Higgs-FSMT})~\cite{de2013anatomy}, and Higgs\_Friendship-Retweet (\textbf{Higgs-FSRT})~\cite{de2013anatomy}.
Basic statistical information of these datasets is displayed in Table~\ref{tab:datasets}.

\textbf{Baselines.} Since no one has studied the node injection attack method for network alignment task before, we designed six heuristic attacking methods and modified three NMA methods as baselines for the comparison. In these baselines, Random attacking method randomly adds intra-links from the injected nodes to the existing nodes to generate perturbed networks. Uniform adds intra-links one by one from the injected nodes to the existing nodes until the budget is used ups. ALDN (prioritizing the attack on large degree nodes) adds intra-links from the injected nodes to the existing nodes based on the descend order of the degree of the existing nodes. ASDN (prioritizing the attack on small degree nodes) adds intra-links from the injected nodes to the existing nodes based on the ascend order of the degree of the existing nodes. AMN (attacking matched nodes only) adds intra-links from the injected nodes to the existing nodes which have been matched. AUMN (attacking unmatched nodes only) adds intra-links from the injected nodes to the existing nodes which have not been matched. LPS~\cite{TANG2022109095} is a NMA method. It uses a weighted random walk algorithm to select the intra-links which have the more significant impact on the network alignment. Then, it removes or switches these selected links. We modified LPS to make it compatible with the GIA scenario. In the modified version, the new intra-links are added from the injected nodes to the existing nodes that the selected links are connected. GPS is the other NMA method proposed in Ref.~\cite{TANG2022109095} which calculates the weights via the global information of the given network to select intra-links. We take the similar way with LPS to modify GPS. EDA~\cite{YuShanqing2021} used an evolutionary approach to perturb link prediction. We modified it by adding intra-links from the injected nodes and the selected exist nodes.

\textbf{SNA methods.} To comprehensively evaluate the efficacy of attack methods, we employed SOIDP~\cite{ding2021soidp}, IDP~\cite{tang2020interlayer}, IONE~\cite{LiuLi2016}, FRUI~\cite{ZhouXiaoping2016}, and CN~\cite{LvLinyuan2011} as the target network alignment approaches for our attacks, assessing their network alignment performance before and after being subjected to attacks by both DPNIA and other baseline methods. It is evident that an effective attack method should induce disruptive effects across various SNA techniques. As known, methods for network alignment by using structural information can be categorized into neighborhood-based and embedding-based methods~\cite{ShuKai2017}. The five SNA methods we employed encompass both the neighborhood-based and network embedding-based approaches. Particularly, FRUI and IONE serve as prominent exemplars of neighborhood-based and embedding-based methodologies, respectively, making significant contributions to the advancement and scholarly exploration of these two methodological categories. In addition, as the primary focus of our experiments lies in assessing the interference capabilities of various attack methods on network alignment algorithms or models, we optimized some of the network alignment algorithms to reduce their computational time, enhancing experiment efficiency.

\textbf{Evaluation metrics.}
We adopted multiple widely recognized evaluation metrics in the field of network alignment, including $Precesion@N$($P@N$), $MAP$, $Precision$, $Recall$, $F1$, and $AUC$, to comprehensively assess how the performance of network alignment methods is affected by various attack methods. Among them, $P@N$ is defined as
\begin{equation}
P@N=\sum_{i=1}^m{\mathds{1}_i\{success@N\}/m}.
\end{equation}
The meaning of ${\mathds{1}_i\{success@N\}}$ is whether there exists the correct inter-link in the top-$N$ list for node $u^\alpha_i$.  If there is, ${\mathds{1}_i\{success@N\}}=1$; otherwise, ${\mathds{1}_i\{success@N\}}=0$. In addition, the value of N ranges from 1 to the number of unmatched nodes and $m$ is the number of all unobserved inter-links. $MAP$ is defined as
\begin{equation}
MAP=(\sum_{i=1}^n{\frac{1}{r_i}})/m.
\end{equation}
It is used to evaluate the ranked performance of a network alignment method and $r_i$ in the equation represents the rank of $E^{\alpha\beta}_i$. $Recall$ is defined as
\begin{equation}
Recall=\frac{TP}{TP+FN},
\end{equation}
where TP is the number of true positives, and FN is the number of false negatives. $Precision$ is defined as
\begin{equation}
Precision=\frac{TP}{TP+FP},
\end{equation}
where FP is the number of false positives. $F1$ is defined as
\begin{equation}
F1=\frac{2 \cdot Recall \cdot Precision}{Recall + Precision}.
\end{equation}
$AUC$ is defined as
\begin{equation}
AUC=(\sum_{i=1}^m{\frac{m_n+1-r_i}{m_n}})/m,
\label{eq_auc}
\end{equation}
where $m_n$ is the number of negative candidate pairs. Clearly, a higher value of those metrics means a better performance of the network alignment methods.

\textbf{Other settings.} For each dataset, we randomly split the inter-links into $(90\%)$ labeled links for training procedure and $(10\%)$ unlabeled links as test set to evaluate the baselines and proposed DPNIA method. The value of $\delta$ can be set by a group of experiments. We perform the random split ten times and report the averaged results.

\begin{table*}[!thb]
\renewcommand\arraystretch{0.8}
\centering
\caption{Results of attacking two networks. A smaller value indicates more effective attacking. }
\setlength{\tabcolsep}{3pt}
\begin{tabular}{c|c|cccc|cccc}
\hline
\multicolumn{2}{c|}{\multirow{2}*{}}&\multicolumn{4}{c|}{P@30}& \multicolumn{4}{c}{MAP} \\
\cline{3-10}
\multicolumn{2}{c|}{}&FT1&TF2&FSMT&FSRT&FT1&TF2&FSMT&FSRT\\
\hline
\multirow{11}*{SOIDP}&original&$0.7165$&$0.8424$&$0.7975$&$0.8066$&$0.4283$&$0.5434$&$0.4303$&$0.4597$\\
~&Random&$0.6589$&$0.6591$&$0.5540$&$0.5887$&$0.3814$&$0.4341$&$0.2827$&$0.2913$\\
~&Uniform&$0.6570$&$0.6699$&$0.5783$&$0.6194$&$0.3725$&$0.4692$&$0.3232$&$0.3534$\\
~&ALDN&$0.6529$&$0.5388$&$\uline{0.4545}$&$\uline{0.4310}$&$\uline{0.3468}$&$0.3206$&$0.2268$&$\uline{0.2161}$\\
~&ASDN&$0.6711$&$0.6435$&$0.4828$&$0.4369$&$0.3961$&$0.3785$&$0.2428$&$0.2399$\\
~&AMN&$\uline{0.6248}$&$\uline{0.5233}$&$0.4890$&$0.5024$&$0.3753$&$\uline{0.3049}$&$0.2358$&$0.2577$\\
~&AUMN&$0.7160$&$0.8335$&$0.7377$&$0.7179$&$0.4282$&$0.5408$&$0.4194$&$0.4081$\\
~&GPS&$0.6896$&$0.6244$&$0.6066$&$0.6279$&$0.4193$&$0.4496$&$0.3315$&$0.3718$\\
~&LPS&$0.6351$&$0.5454$&$0.4779$&$0.5340$&$0.3580$&$0.3162$&$\uline{0.2124}$&$0.2637$\\
~&EDA&$-$&$0.6052$&$-$&$-$&$-$&$0.3686$&$-$&$-$\\
~&DPNIA&$\bm{0.4414}$&$\bm{0.1547}$&$\bm{0.3616}$&$\bm{0.3858}$&$\bm{0.1020}$&$\bm{0.0394}$&$\bm{0.0563}$&$\bm{0.0570}$\\
\cline{1-10}
\multirow{11}*{IDP}&original&$0.7136$&$0.8500$&$0.7943$&$0.8039$&$0.4197$&$0.5583$&$0.4203$&$0.4539$\\
~&Random&$0.6643$&$0.6372$&$0.4705$&$0.4345$&$0.3814$&$0.3877$&$0.2292$&$0.2333$\\
~&Uniform&$0.6610$&$0.6271$&$\uline{0.4673}$&$\uline{0.4321}$&$0.3744$&$0.4124$&$\uline{0.2287}$&$\uline{0.2306}$\\
~&ALDN&$0.6613$&$\uline{0.5790}$&$0.4682$&$0.4408$&$\uline{0.3500}$&$0.3541$&$0.2400$&$0.2359$\\
~&ASDN&$0.6770$&$0.6624$&$0.4956$&$0.4541$&$0.3990$&$0.4134$&$0.2571$&$0.2617$\\
~&AMN&$\uline{0.6359}$&$0.6040$&$0.4975$&$0.5191$&$0.3754$&$0.3726$&$0.2385$&$0.2718$\\
~&AUMN&$0.7136$&$0.8500$&$0.6661$&$0.5733$&$0.4197$&$0.5583$&$0.3701$&$0.3482$\\
~&GPS&$0.7024$&$0.7267$&$0.6180$&$0.6449$&$0.4155$&$0.5020$&$0.3288$&$0.3763$\\
~&LPS&$0.6533$&$0.5956$&$0.4920$&$0.5483$&$0.3733$&$\uline{0.3404}$&$0.2360$&$0.2879$\\
~&EDA&$-$&$0.6630$&$-$&$-$&$-$&$0.4273$&$-$&$-$\\
~&DPNIA&$\bm{0.4562}$&$\bm{0.2514}$&$\bm{0.3781}$&$\bm{0.4074}$&$\bm{0.1200}$&$\bm{0.1260}$&$\bm{0.0961}$&$\bm{0.0947}$\\
\cline{1-10}
\multirow{11}*{IONE}&original&$0.6575$&$0.8648$&$0.7364$&$0.7320$&$0.4147$&$0.6040$&$0.3698$&$0.4210$\\
~&Random&$0.6537$&$0.9818$&$0.9397$&$0.9304$&$0.4124$&$0.7598$&$0.5932$&$0.5587$\\
~&Uniform&$0.6702$&$0.9937$&$0.9627$&$0.9652$&$0.4224$&$0.8360$&$0.6663$&$0.6416$\\
~&ALDN&$0.6573$&$0.8517$&$0.6965$&$\uline{0.6615}$&$0.4170$&$0.5942$&$0.3488$&$\bm{0.3465}$\\
~&ASDN&$0.6506$&$0.8545$&$0.7075$&$0.6712$&$0.4271$&$0.6049$&$0.3504$&$\uline{0.3556}$\\
~&AMN&$0.6440$&$0.8692$&$0.7322$&$0.7234$&$0.3909$&$0.5868$&$0.3592$&$0.3942$\\
~&AUMN&$\uline{0.6207}$&$\bm{0.7507}$&$0.8875$&$0.8841$&$0.3815$&$\uline{0.5078}$&$0.4784$&$0.4872$\\
~&GPS&$0.6390$&$0.8275$&$\uline{0.6836}$&$0.6693$&$\uline{0.3811}$&$0.5352$&$\bm{0.3211}$&$0.3622$\\
~&LPS&$0.6522$&$0.8531$&$0.7117$&$0.7152$&$0.3970$&$0.5348$&$0.3484$&$0.3923$\\
~&EDA&$-$&$0.8399$&$-$&$-$&$-$&$\uline{0.5078}$&$-$&$-$\\
~&DPNIA&$\bm{0.6047}$&$\uline{0.7747}$&$\bm{0.6486}$&$\bm{0.6374}$&$\bm{0.3581}$&$\bm{0.4925}$&$\uline{0.3372}$&$0.3670$\\
\cline{1-10}
\multirow{11}*{FRUI}&original&$0.6778$&$0.8504$&$0.7676$&$0.7957$&$0.3887$&$0.5266$&$0.3858$&$0.4328$\\
~&Random&$0.6495$&$0.6882$&$0.4972$&$0.4679$&$0.3670$&$0.4256$&$0.2254$&$0.2380$\\
~&Uniform&$0.6473$&$0.6861$&$0.4958$&$\uline{0.4667}$&$0.3603$&$0.4359$&$\uline{0.2236}$&$0.2361$\\
~&ALDN&$0.6443$&$\uline{0.6192}$&$0.4978$&$0.4726$&$\uline{0.3226}$&$0.3795$&$0.2345$&$\uline{0.2350}$\\
~&ASDN&$0.6633$&$0.7161$&$0.5354$&$0.4881$&$0.3800$&$0.4447$&$0.2613$&$0.2730$\\
~&AMN&$0.6363$&$0.6668$&$0.5223$&$0.5569$&$0.3707$&$0.4026$&$0.2319$&$0.2752$\\
~&AUMN&$0.6759$&$0.8476$&$0.6350$&$0.5595$&$0.3872$&$0.5329$&$0.3251$&$0.3188$\\
~&GPS&$0.6703$&$0.7174$&$0.5712$&$0.6296$&$0.3900$&$0.4803$&$0.2946$&$0.3559$\\
~&LPS&$\uline{0.6313}$&$0.6248$&$\uline{0.4814}$&$0.5654$&$0.3610$&$0.3636$&$0.2255$&$0.2906$\\
~&EDA&$-$&$0.7101$&$-$&$-$&$-$&$0.4540$&$-$&$-$\\
~&DPNIA&$\bm{0.4254}$&$\bm{0.2639}$&$\bm{0.3744}$&$\bm{0.4128}$&$\bm{0.1127}$&$\bm{0.1269}$&$\bm{0.0953}$&$\bm{0.0950}$\\
\cline{1-10}
\multirow{11}*{CN}&original&$0.6401$&$0.8385$&$0.7536$&$0.7836$&$0.3593$&$0.4964$&$0.3534$&$0.3987$\\
~&Random&$0.6208$&$0.6811$&$0.4890$&$0.4636$&$0.3439$&$0.4086$&$0.2219$&$0.2340$\\
~&Uniform&$0.6185$&$0.6768$&$0.4856$&$\uline{0.4601}$&$0.3382$&$0.4266$&$\uline{0.2204}$&$0.2320$\\
~&ALDN&$0.6151$&$\uline{0.6130}$&$0.4901$&$0.4669$&$\uline{0.3043}$&$0.3688$&$0.2295$&$\uline{0.2310}$\\
~&ASDN&$0.6309$&$0.7085$&$0.5233$&$0.4830$&$0.3534$&$0.4289$&$0.2534$&$0.2666$\\
~&AMN&$0.6109$&$0.6608$&$0.5142$&$0.5517$&$0.3480$&$0.3938$&$0.2292$&$0.2711$\\
~&AUMN&$0.6401$&$0.8385$&$0.6227$&$0.5533$&$0.3593$&$0.4964$&$0.3066$&$0.3034$\\
~&GPS&$0.6366$&$0.7164$&$0.5659$&$0.6235$&$0.3585$&$0.4621$&$0.2749$&$0.3322$\\
~&LPS&$\uline{0.6100}$&$0.6194$&$\uline{0.4787}$&$0.5598$&$0.3417$&$\uline{0.3514}$&$0.2207$&$0.2817$\\
~&EDA&$-$&$0.7047$&$-$&$-$&$-$&$0.4321$&$-$&$-$\\
~&DPNIA&$\bm{0.4175}$&$\bm{0.2615}$&$\bm{0.3724}$&$\bm{0.4114}$&$\bm{0.1093}$&$\bm{0.1261}$&$\bm{0.0950}$&$\bm{0.0948}$\\
\cline{1-10}
\end{tabular}
\label{tab:datasets_twolys}
\end{table*}

\subsection{Attacking multiple networks simultaneously}
We initiated our comparison by evaluating the performance of DPNIA in comparison to nine baselines under the condition of attacking multiple networks simultaneously. In terms of the attack budget, we configured the number of injected nodes to be 200, with an average node degree of 200. Results under different budgets are compared in subsequent experiments. The experimental results are presented in Table~\ref{tab:datasets_twolys}. The most effective attack methods have been denoted in bold black text, and the second-best attack methods have been marked with an underline beneath the metric values. It is worthwhile note that EDA has not generated any results for the FT1, FSMT, and FSRT datasets. The reason for this issue is that among the three datasets, each of them contains a social network with over one hundred thousand intra-links. EDA has been running on our server for several months without yielding any outcomes.

As shown in the table, the application of DPNIA in attack scenarios yielded significant results, notably resulting in an average reduction of $34.9\%$ and a maximum reduction of $68.8\%$ in the $P@30$ metric for the network alignment algorithms when compared to their performance in an unattacked setting. Meanwhile, for the metric of MAP, DPNIA resulting in an average reduction of $28.8\%$ and a maximum reduction of $50.4\%$. Compared to the best baselines under various alignments and datasets, DPNIA exhibited an average reduction of $13.6\%$ and a maximum reduction of $36.9\%$ in the $P@30$ metric. In terms of the MAP metric, these reductions amounted to $14.7\%$ and $26.5\%$, respectively. While employing IONE as the network alignment algorithm, AUMN takes the lowest $P@30$ value on the TF2 dataset, GPS takes the lowest MAP value on the FSMT dataset, and ALDN takes the lowest MAP value on the TFRT datasets. However, it is crucial to emphasize that these attack methods exhibited less effectiveness under the other conditions and the differences between DPNIA and those best attack methods were relatively modest. The observations above indicate that the method presented in this paper exhibits an enhanced capability to attack the existed network alignment algorithms or models.

Table~\ref{tab:datasets_twolys} also presents the following observations. Firstly, none of the baseline methods consistently outperformed the others across the majority of conditions. In a few specific instances, ALDN, AMN, GPS, LPS, and others have emerged as the second-best performers, surpassing all other baseline methods. This, to some extent, provides evidence illustrating the effectiveness of our proposed method, which consistently shows the highest attack capabilities across nearly all conditions, while other methods struggle to consistently achieve even the second-best attack performance.

Secondly, the order of metric values for ALDN and ASDN corresponds closely to the order of metric values for LPS and GPS. For instance, in most cases, ALDN reports lower $P@30$ and MAP values compared to ASDN, similar to how LPS's metrics are lower than GPS's. This correlation can be attributed to the fact that ALDN prioritizes establishing intra-links from the injected nodes to existing nodes with higher degrees, similar to how LPS tends to favor nodes with larger degrees when compared to GPS.
\begin{table*}[!t]
\renewcommand\arraystretch{0.8}
\centering
\caption{Results of after attacking one network.}
\setlength{\tabcolsep}{3pt}
\begin{tabular}{c|c|cccc|cccc}

\hline
\multicolumn{2}{c|}{\multirow{2}*{}}&\multicolumn{4}{c|}{P@30}& \multicolumn{4}{c}{MAP} \\
\cline{3-10}
\multicolumn{2}{c|}{}&FT1&TF2&FSMT&FSRT&FT1&TF2&FSMT&FSRT\\
\hline
\multirow{11}*{SOIDP}&original&$0.7165$&$0.8424$&$0.7975$&$0.8066$&$0.4283$&$0.5434$&$0.4303$&$0.4597$\\
~&Random&$0.6996$&$0.7246$&$0.7750$&$0.7819$&$0.4202$&$0.4486$&$0.4094$&$0.4445$\\
~&Uniform&$0.6991$&$0.7170$&$0.7737$&$0.7830$&$0.4187$&$0.4596$&$0.4112$&$0.4461$\\
~&ALDN&$0.6996$&$0.6964$&$\uline{0.7700}$&$0.7805$&$0.4204$&$0.4400$&$0.4143$&$0.4500$\\
~&ASDN&$0.7011$&$0.7394$&$0.7767$&$0.7830$&$0.4221$&$0.4574$&$0.4167$&$0.4508$\\
~&AMN&$\uline{0.6880}$&$0.7011$&$0.7722$&$\uline{0.7737}$&$0.4154$&$\uline{0.4308}$&$0.4050$&$\uline{0.4236}$\\
~&AUMN&$0.7163$&$0.8383$&$0.7935$&$0.8017$&$0.4282$&$0.5423$&$0.4279$&$0.4689$\\
~&GPS&$0.7081$&$0.7591$&$0.7863$&$0.7972$&$0.4253$&$0.5124$&$0.4266$&$0.4564$\\
~&LPS&$0.6967$&$\uline{0.6962}$&$0.7725$&$0.7843$&$\uline{0.4117}$&$0.4329$&$\uline{0.4024}$&$0.4322$\\
~&EDA&$-$&$0.7372$&$-$&$-$&$-$&$0.4645$&$-$&$-$\\
~&DPNIA&$\bm{0.6454}$&$\bm{0.5053}$&$\bm{0.7156}$&$\bm{0.7330}$&$\bm{0.2922}$&$\bm{0.2985}$&$\bm{0.2571}$&$\bm{0.2802}$\\
\cline{1-10}
\multirow{11}*{IDP}&original&$0.7136$&$0.8500$&$0.7943$&$0.8039$&$0.4197$&$0.5583$&$0.4203$&$0.4539$\\
~&Random&$0.6994$&$0.7504$&$0.7778$&$0.7870$&$0.4147$&$0.4827$&$0.4081$&$0.4509$\\
~&Uniform&$0.6991$&$0.7506$&$0.7773$&$0.7864$&$0.4133$&$0.4975$&$0.4090$&$0.4519$\\
~&ALDN&$0.6989$&$\uline{0.7281}$&$\uline{0.7737}$&$0.7854$&$0.4144$&$0.4705$&$0.4113$&$0.4528$\\
~&ASDN&$0.7003$&$0.7613$&$0.7784$&$0.7877$&$0.4156$&$0.4890$&$0.4129$&$0.4545$\\
~&AMN&$\uline{0.6905}$&$0.7363$&$0.7777$&$\uline{0.7816}$&$\uline{0.4108}$&$0.4742$&$0.4064$&$\uline{0.4354}$\\
~&AUMN&$0.7136$&$0.8500$&$0.7905$&$0.8016$&$0.4197$&$0.5583$&$0.4182$&$0.4635$\\
~&GPS&$0.7110$&$0.7957$&$0.7885$&$0.7985$&$0.4188$&$0.5365$&$0.4191$&$0.4519$\\
~&LPS&$0.7003$&$0.7300$&$0.7772$&$0.7892$&$0.4121$&$\uline{0.4563}$&$\uline{0.4034}$&$0.4368$\\
~&EDA&$-$&$0.7641$&$-$&$-$&$-$&$0.4978$&$-$&$-$\\
~&DPNIA&$\bm{0.6533}$&$\bm{0.5567}$&$\bm{0.7254}$&$\bm{0.7482}$&$\bm{0.3046}$&$\bm{0.3476}$&$\bm{0.2924}$&$\bm{0.3140}$\\
\cline{1-10}
\multirow{11}*{IONE}&original&$0.6522$&$0.8404$&$0.7044$&$0.6926$&$0.4125$&$0.5591$&$0.3566$&$0.4036$\\
~&Random&$0.6685$&$0.8831$&$0.6745$&$0.6726$&$0.4131$&$0.6175$&$0.3451$&$0.3793$\\
~&Uniform&$0.6279$&$0.8560$&$\uline{0.6645}$&$\uline{0.6654}$&$0.3810$&$0.5755$&$0.3189$&$0.3666$\\
~&ALDN&$0.6901$&$0.8788$&$0.6938$&$0.6883$&$0.4245$&$0.6197$&$0.3605$&$0.3906$\\
~&ASDN&$0.6841$&$0.8799$&$0.6999$&$0.6881$&$0.4286$&$0.6175$&$0.3599$&$0.3942$\\
~&AMN&$0.6866$&$0.8911$&$0.7054$&$0.6853$&$0.4202$&$0.6385$&$0.3506$&$0.3804$\\
~&AUMN&$0.6482$&$\bm{0.7594}$&$0.6768$&$0.6772$&$0.3916$&$\bm{0.5116}$&$0.3267$&$0.3721$\\
~&GPS&$\uline{0.6253}$&$0.8509$&$0.6755$&$0.6742$&$\uline{0.3746}$&$0.5409$&$\uline{0.3138}$&$\uline{0.3559}$\\
~&LPS&$0.6394$&$0.8553$&$0.6942$&$0.6822$&$0.3859$&$0.5344$&$0.3304$&$0.3625$\\
~&EDA&$-$&$0.8754$&$-$&$-$&$-$&$0.6009$&$-$&$-$\\
~&DPNIA&$\bm{0.6189}$&$\uline{0.8275}$&$\bm{0.6527}$&$\bm{0.6526}$&$\bm{0.3681}$&$\uline{0.5302}$&$\bm{0.3066}$&$\bm{0.3438}$\\
\cline{1-10}
\multirow{11}*{FRUI}&original&$0.6778$&$0.8504$&$0.7676$&$0.7957$&$0.3887$&$0.5266$&$0.3858$&$0.4328$\\
~&Random&$0.6702$&$0.7770$&$0.7611$&$0.7868$&$0.3860$&$0.4987$&$0.3828$&$0.4413$\\
~&Uniform&$0.6681$&$0.7752$&$0.7607$&$0.7862$&$0.3862$&$0.5074$&$0.3828$&$0.4417$\\
~&ALDN&$\uline{0.6674}$&$0.7472$&$\uline{0.7578}$&$\uline{0.7857}$&$0.3855$&$0.4805$&$0.3826$&$0.4413$\\
~&ASDN&$0.6704$&$0.7884$&$0.7608$&$0.7879$&$0.3866$&$0.5044$&$0.3835$&$0.4423$\\
~&AMN&$0.6711$&$0.7709$&$0.7636$&$0.7866$&$0.3858$&$0.4743$&$0.3832$&$0.4287$\\
~&AUMN&$0.6690$&$0.8452$&$0.7631$&$0.7914$&$0.3867$&$0.5344$&$0.3843$&$0.4432$\\
~&GPS&$0.6738$&$0.7906$&$0.7658$&$0.7913$&$0.3876$&$0.5199$&$0.3853$&$0.4320$\\
~&LPS&$0.6683$&$\uline{0.7443}$&$0.7613$&$0.7883$&$\uline{0.3849}$&$\uline{0.4680}$&$\uline{0.3822}$&$\uline{0.4282}$\\
~&EDA&$-$&$0.7910$&$-$&$-$&$-$&$0.5087$&$-$&$-$\\
~&DPNIA&$\bm{0.6265}$&$\bm{0.5484}$&$\bm{0.7116}$&$\bm{0.7452}$&$\bm{0.2864}$&$\bm{0.3382}$&$\bm{0.2860}$&$\bm{0.3202}$\\
\cline{1-10}
\multirow{11}*{CN}&original&$0.6401$&$0.8385$&$0.7536$&$0.7836$&$0.3593$&$0.4964$&$0.3534$&$0.3987$\\
~&Random&$0.6382$&$0.7699$&$0.7478$&$0.7771$&$0.3586$&$0.4615$&$0.3500$&$0.4074$\\
~&Uniform&$0.6371$&$0.7685$&$0.7474$&$0.7766$&$0.3583$&$0.4704$&$0.3500$&$0.4079$\\
~&ALDN&$0.6361$&$0.7405$&$\uline{0.7445}$&$0.7760$&$0.3584$&$0.4435$&$0.3498$&$0.4075$\\
~&ASDN&$0.6390$&$0.7814$&$0.7477$&$0.7783$&$0.3589$&$0.4668$&$0.3507$&$0.4084$\\
~&AMN&$\uline{0.6352}$&$0.7606$&$0.7499$&$\uline{0.7743}$&$0.3583$&$0.4541$&$0.3510$&$0.3953$\\
~&AUMN&$0.6401$&$0.8385$&$0.7498$&$0.7818$&$0.3593$&$0.4964$&$0.3515$&$0.4094$\\
~&GPS&$0.6383$&$0.7851$&$0.7523$&$0.7802$&$0.3589$&$0.4843$&$0.3530$&$0.3983$\\
~&LPS&$0.6368$&$\uline{0.7366}$&$0.7484$&$0.7780$&$\uline{0.3577}$&$\uline{0.4333}$&$\uline{0.3495}$&$\uline{0.3942}$\\
~&EDA&$-$&$0.7846$&$-$&$-$&$-$&$0.4717$&$-$&$-$\\
~&DPNIA&$\bm{0.5989}$&$\bm{0.5421}$&$\bm{0.6982}$&$\bm{0.7357}$&$\bm{0.2642}$&$\bm{0.3237}$&$\bm{0.2538}$&$\bm{0.2868}$\\
\cline{1-10}
\end{tabular}
\label{tab_datasets_lay2_200}
\end{table*}

Finally, the utilization of Uniform and Random unexpectedly resulted in improvements in both $P@30$ and MAP metrics in many cases when employing IONE as the network alignment method. We also observed this phenomenon in experiments conducted under additional conditions. The explanation may lie in that the Uniform method adds intra-links one by one from the injected nodes to the existing nodes. As a result, low-degree nodes acquire improved vector representations, facilitating a more accurate alignment with their corresponding nodes. For the high-degree nodes, the influence on their node representations is comparatively negligible. It is well-established that many online social networks exhibit scale-free degree distributions~\cite{ZhouTao2011}, where the majority of nodes have small degrees, while only a few nodes have large degrees~\cite{tang2020interlayer}. The same reason can be applied to explain the Random method, which iteratively traverses existing nodes one by one and then, based on the set probability, determines whether to establish a intra-links between the currently traversed node and the injected node. Therefore, we can observed that its $P@30$ and MAP metrics are lower than those of the Uniform method. Similar phenomena occasionally occur in other baselines, attributed to similarities with the Uniform and Random methods. However, due to their less pronounced influence on node embeddings, their $P@30$ and MAP values are lower than those of the Uniform and Random methods. The observed phenomenon, to some extent, corroborates the effectiveness of the method proposed in this paper. It indicates that injecting nodes does not necessarily result in a decrease in the accuracy of network alignment. The effective attack on network alignment algorithms or models through node injection requires systematic analysis and careful design.
\begin{figure} [t!]
    \centering
        \subfigure[Average $P@30$.]{
        \begin{minipage}[t]{0.48\linewidth}
        \centering
        \includegraphics[width=4.5cm]{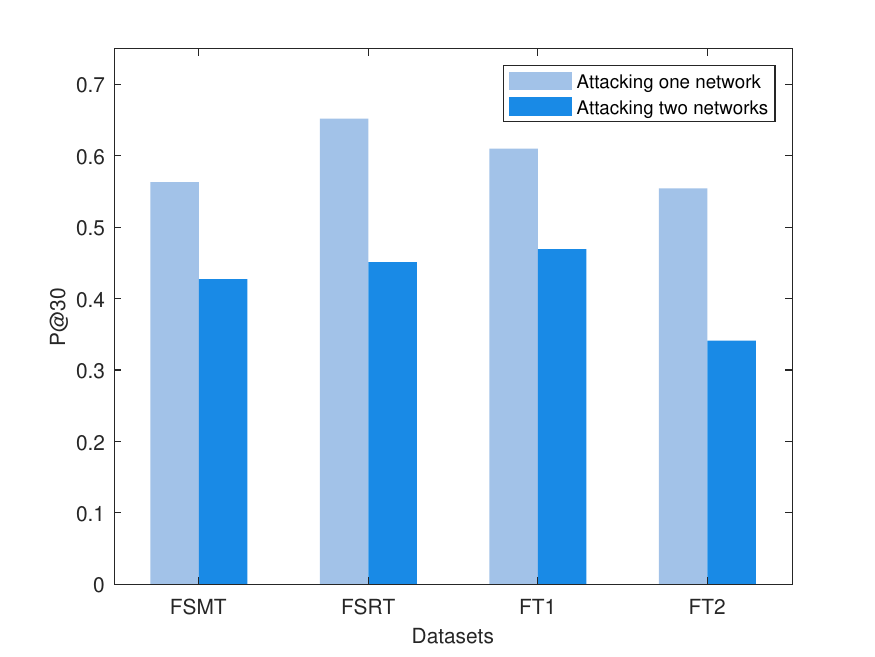}
        \end{minipage}%
    }%
        \subfigure[Average $MAP$.]{
        \begin{minipage}[t]{0.48\linewidth}
        \centering
        \includegraphics[width=4.5cm]{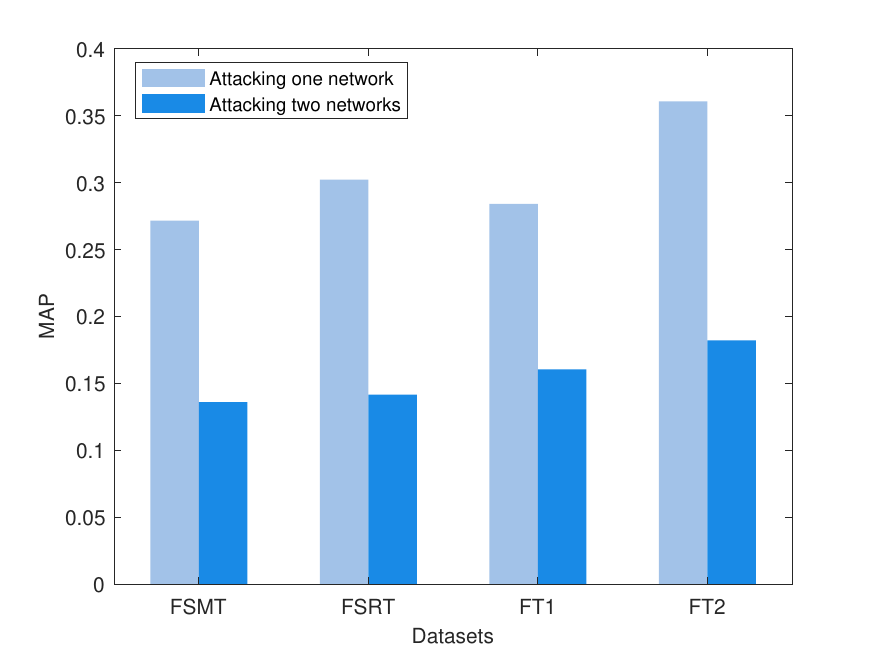}
        \end{minipage}%
    }%
    \caption{Comparison between the scenarios of attacking only one network and attacking two networks.}
    \label{pic_compdifflaynum}
\end{figure}

\subsection{Attacking one network}
We also conducted experiments on comparing DPNIA with the nine baselins under the condition of attacking single network in each dataset. In terms of the attack budget, we configured the number of injected nodes to be 200, with an average node degree of 200. The experimental results are presented in Table~\ref{tab_datasets_lay2_200}.

As shown in the table, the application of DPNIA in attack scenarios yielded significant results, notably resulting in an average reduction of $10.4\%$ and a maximum reduction of $33.7\%$ in the $P@30$ metric for the network alignment algorithms when compared to their performance in an unattacked setting. Meanwhile, for the metric of MAP, DPNIA resulting in an average reduction of $12.5\%$ and a maximum reduction of $24.5\%$. Compared to the best baselines under various alignments and datasets, DPNIA exhibited an average reduction of $6.1\%$ and a maximum reduction of $19.6\%$ in the $P@30$ metric. In terms of the MAP metric, these reductions amounted to $9.2\%$ and $14.5\%$, respectively. While employing IONE as the network alignment algorithm, AUMN takes the lowest $P@30$ and MAP values on the TF2 dataset. However, it is crucial to emphasize that this attack method exhibited less effectiveness under the other conditions and the differences between DPNIA and that attack method were relatively modest. The observations above restate that the method presented in this paper exhibits an enhanced capability to attack the existed network alignment algorithms or models.

Table~\ref{tab_datasets_lay2_200} also presents the following observations. When attacking only one network, similarly, none of the baseline methods consistently outperformed the others across the majority of conditions. Meanwhile, the observation that the order of metric values for ALDN and ASDN corresponds closely to the order of metric values for LPS and GPS in Table~\ref{tab:datasets_twolys} can still be observed in Table~\ref{tab_datasets_lay2_200}. Additionally, the unexpected improvement in both the $P@30$ and MAP metrics when employing IONE as the network alignment method, as observed in Table~\ref{tab:datasets_twolys}, has been observed in Table~\ref{tab_datasets_lay2_200} again. Nevertheless, this improvement is less pronounced than that observed in Table~\ref{tab:datasets_twolys}, as the attack was restricted to a single network within the dataset.
\begin{figure} [t!]
    \centering
        \subfigure[Average $P@30$]{
        \begin{minipage}[t]{0.48\linewidth}
        \centering
        \includegraphics[width=4.5cm]{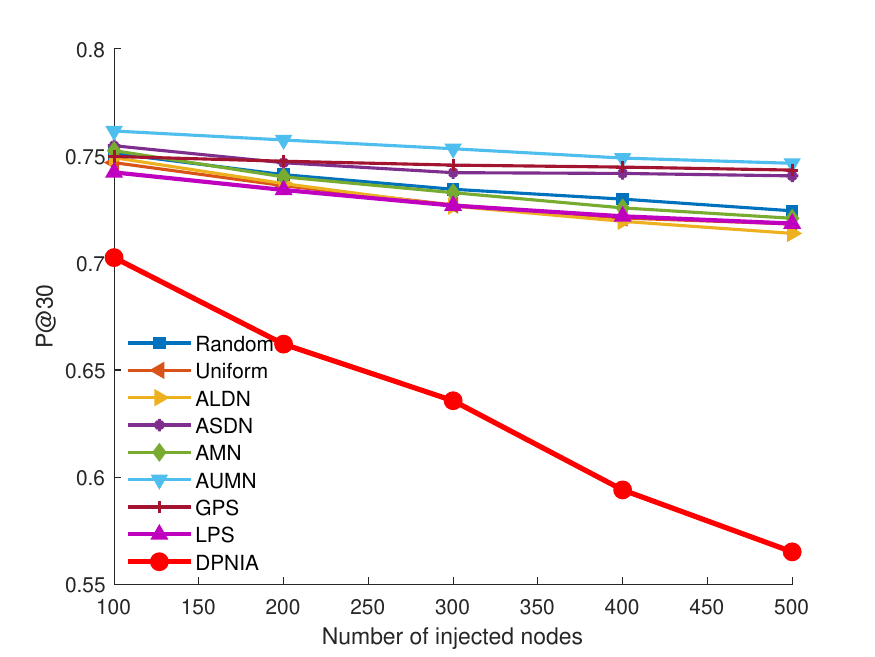}
        \end{minipage}%
    }%
        \subfigure[Average $MAP$]{
        \begin{minipage}[t]{0.48\linewidth}
        \centering
        \includegraphics[width=4.5cm]{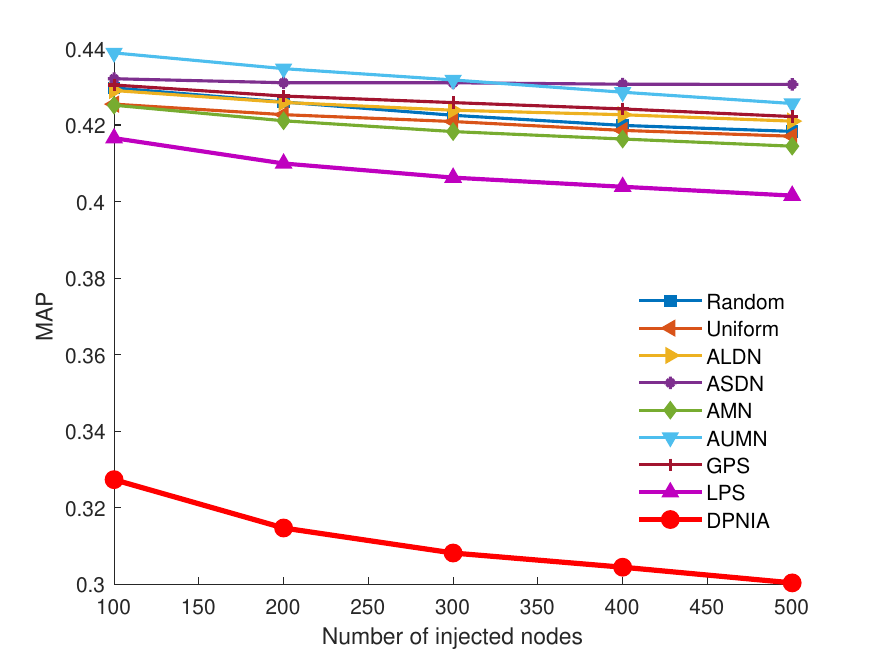}
        \end{minipage}%
    }
    \centering

    \caption{Network alignment performance on the four datasets by varying the number of injected nodes.}
    \label{pic_diffnodenum_avg}
\end{figure}

We have compared the scenarios of attacking only one network and attacking two networks. The results are shown in Figure~\ref{pic_compdifflaynum}. To ensure a fair comparison, when attacking a single network, we set the number of injected nodes to 400. As depicted in the figure, under the same budget constraints, attacking two networks has a more significant detrimental impact on the performance of SNA. For instance, on the FSRT dataset, after applying DPNIA to attack two networks, the average $P@30$ of the five network alignment methods is $0.4510$, whereas the $P@30$ after attacking one network is $0.6516$. Attacking two networks results in a decrease of $0.2006$ in the average $P@30$ compared to attacking one network.

\subsection{Effect of number of injected nodes}
As we discussed that the budget $\Delta$ is essential to the attack, we investigate the performance of the attacking methods by varying the injected nodes number, setting to 100, 200, 300, 400, and 500. The average node degree of the injected nodes is set to 200. In this group of experiments, we performed the attack on a single network for each dataset.

Figure~\ref{pic_diffnodenum_avg} depicts the average $P@30$ and $MAP$ of the five network alignment methods under different attacks varing numbers of injected nodes. We can observe that the proposed DPNIA exhibits the strongest attack capability. Irrespective of the number of injected nodes, it consistently results in the lowest performance of the SNA methods when compared to the baselines. Among these baselines, LPS performs the most effectively. This could be attributed to LPS's inclination to select nodes that have a greater impact on network alignment. It's worth noting that GPS also tends to select nodes that significantly influence network alignment, but its attack capability is slightly inferior to LPS. This could be due to GPS's preference for linking injected nodes to existing low-degree nodes in the network more frequently than LPS does~\cite{TANG2022109095}. In most cases, prioritizing intra-links to existed nodes with high degrees for injected nodes is likely to result in better attack effectiveness than prioritizing intra-links to existed nodes with low degrees. The results from ALDN and ASDN support this conclusion. The corresponding $P@30$ and $MAP$ values for ALDN are significantly lower than those for ASDN, indicating that ALDN outperforms ASDN in terms of attack effectiveness. As mentioned earlier in the baselines subsection, ALDN prioritizes linking injected nodes to existing high-degree nodes, while ASDN prioritizes linking them to low-degree nodes.
\begin{figure*} [t!]
    \centering
    \includegraphics[width=0.99\textwidth]{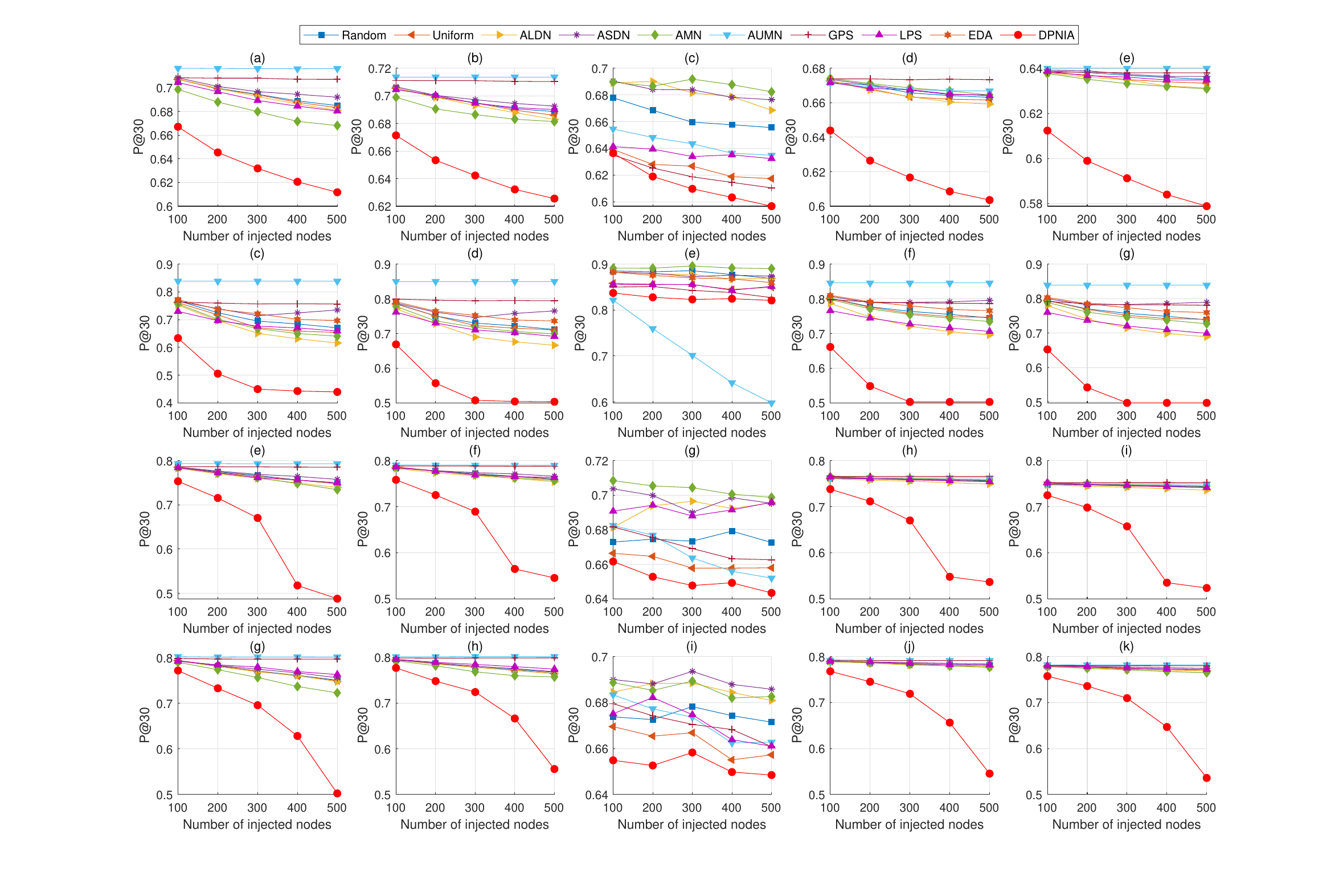}
    \caption{$P@30$ at different numbers of injected nodes. Each row of this figure utilizes the same dataset, arranged from top to bottom as FT1, FT2, FSMT, and FSRT. Each column utilizes the same network alignment method, ordered from left to right as SOIDP, IDP, IONE, FRUI, and CN.}
    \label{pic_p30_diffNodeNum}
\end{figure*}

In most cases with varying numbers of injected nodes, AUMN consistently exhibits the highest values for $P@30$ and $MAP$, suggesting that its attack effectiveness is comparatively the weakest. AMN performs somewhat better than AUMN, indicating that when conducting injection attacks, linking injected nodes to already matched nodes yields superior attack results compared to linking them to unmatched nodes. The results from Random, which falls between AMN and AUMN concerning $P@30$ and $MAP$, further support this conclusion. This is because both AMN and AUMN are fundamentally random methods. AMN randomly adds intra-links between matched nodes to the injected nodes, while AUMN randomly adds intra-links between unmatched nodes to the injected nodes.

To achieve a comprehensive understanding of how various network alignment methods are affected by attacks on different datasets, we presented the details of the $P@30$ in Figure~\ref{pic_p30_diffNodeNum}. Noting that each row of this figure utilizes the same dataset while each column utilizes the same network alignment method. We can see that under the attack of DPNIA, almost all network alignment methods reached their lowest values of $P@30$ for a given number of injected nodes. For instance, on the FSMT dataset when the injected node number is 500, after DPNIA's attack, the $P@30$ for SOIDP is $0.4876$, while the best-performing baseline, AMN, achieves $0.7346$. DPNIA results in a drop of $0.2470$. This observation underscores the effective performance of our proposed DPNIA in the context of attacking network alignment. It's worth noting that in the FT2 dataset, AUMN achieves the best attack performance when IONE is used as the network alignment method. This might be attributed to the low number of unmatched nodes in the FT2 dataset, leading to a significant impact of AUMN on the embedding of unmatched nodes by IONE. However, this seems to be an exceptional case, as AUMN's attacking performance deteriorates obviously when applied to other network alignment methods or datasets. As the number of unmatched nodes increases, even in the case of the FT2 dataset, AUMN would becomes one of the less effective attack methods.

For the same network alignment method within a dataset, increasing the number of injected nodes results in a smaller $P@30$ value. This aligns with our intuition that a larger attack budget leads to more substantial attacking performance. Apart from DPNIA, none of the attacking methods consistently outperformed the others across the majority of conditions. For any baseline attack method, it may yield favorable results under some conditions, but it can also perform poorly under the other conditions. This, to some extent, once again provides evidence illustrating the effectiveness of our proposed method, which consistently shows the highest attack capabilities across nearly all conditions, while other methods struggle to consistently achieve even the second-best attack performance. Figure~\ref{pic_MAP_diffNodeNum}  in Appendix A is the results of $MAP$. As depicted from the figure, the impact of varying the number of injected nodes on the $MAP$ metric is quite consistent with that of $P@30$.
\begin{figure} [t!]
    \centering
        \subfigure[The difference of $P@30$]{
        \begin{minipage}[t]{0.48\linewidth}
        \centering
        \includegraphics[width=4.1cm]{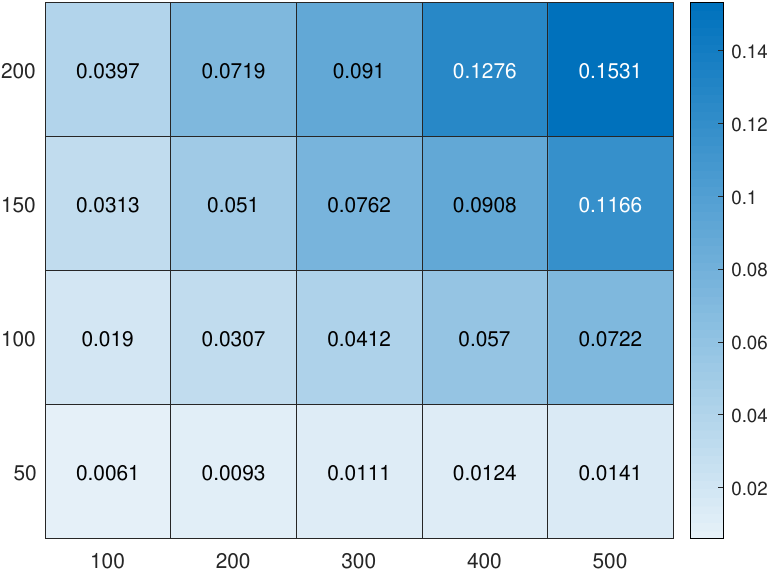}
        \end{minipage}%
    }%
        \subfigure[The difference of $MAP$]{
        \begin{minipage}[t]{0.48\linewidth}
        \centering
        \includegraphics[width=4.1cm]{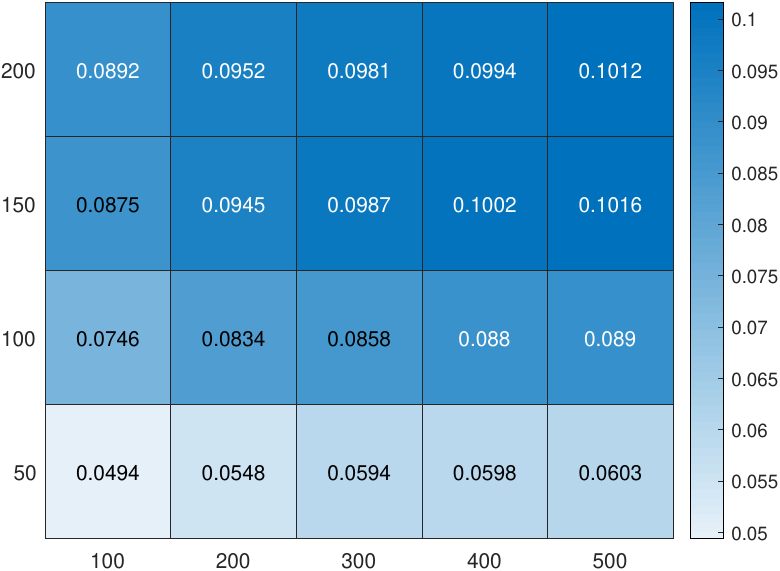}
        \end{minipage}%
    }
    \centering

    \caption{Comparison between DPNIA and LPS.}
    \label{pic_compLPSandDPNIA}
\end{figure}
\subsection{Effect of average node degree}
In addition to the number of injected nodes, the degree of injected nodes, which refers to the number of connections each injected node possesses, is also correlated with the budget. Evidently, as the average degree of injected nodes increases, so does the budget allocation. We investigated the performance of the different network alignment approaches by varying the average node degree of the injected nodes, setting to 50, 100, 150, and 200. The injected node number is set to 200. In this group of experiments, we performed the attack on a single network of each dataset too.

\begin{figure} [t!]
    \centering
        \subfigure[Average $P@30$]{
        \begin{minipage}[t]{0.48\linewidth}
        \centering
        \includegraphics[width=4.5cm]{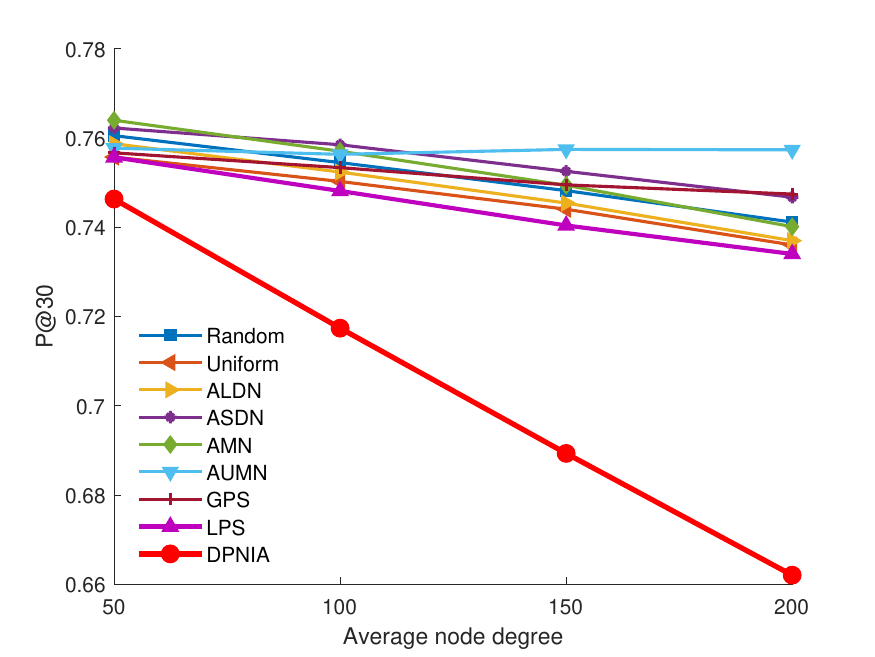}
        \end{minipage}%
    }%
        \subfigure[Average $MAP$]{
        \begin{minipage}[t]{0.48\linewidth}
        \centering
        \includegraphics[width=4.5cm]{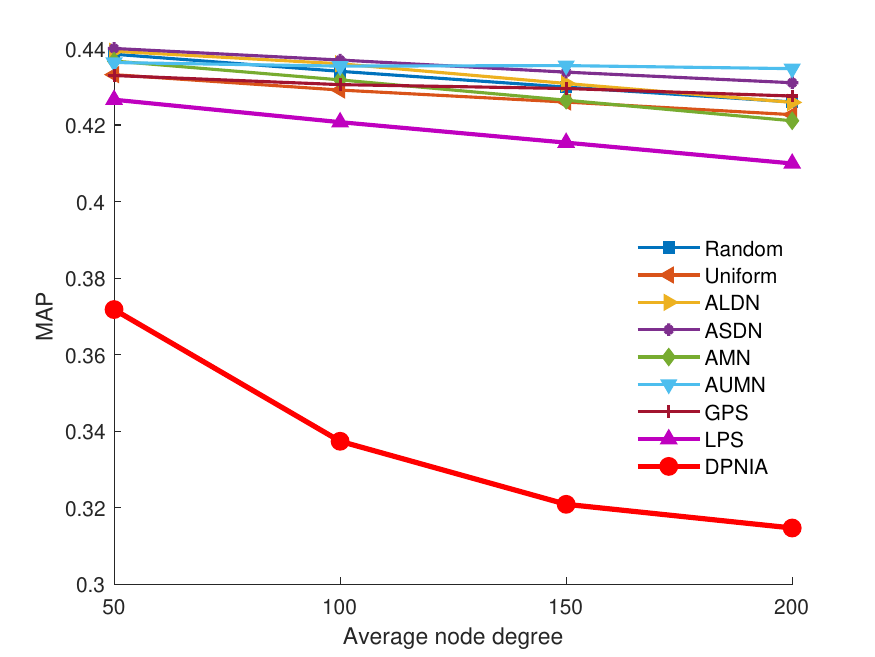}
        \end{minipage}%
    }
    \centering
    \caption{Network alignment performance on the four datasets by varying the average node degree of injected nodes.}
    \label{pic_difflinknum_avg}
\end{figure}

Figure~\ref{pic_difflinknum_avg} depicts the average $P@30$ and $MAP$ of the five network alignment methods under different attacks varing the average node degree of injected nodes. We can see that the average $P@30$ and $MAP$ of our proposed DPNIA are lowest in each subfigure. This observation reveal that DPNIA exhibits the strongest attack capability. Irrespective of the average node degree of injected nodes, it consistently results in the lowest performance of the SNA methods in comparison to the baselines. LPS performs the most effectively when compared to the other baselines. GPS's attack capability is also slightly inferior to LPS. Meanwhile, the corresponding $P@30$ and $MAP$ values for ALDN are significantly lower than those for ASDN. Additionly, in most cases with varying average node degree of injected nodes, AUMN consistently records the highest values for $P@30$ and $MAP$, whereas AMN performs marginally better than AUMN. Random also falls between AMN and AUMN. These observations can further corroborate the conclusions we derived in Figure~\ref{pic_diffnodenum_avg}.

\begin{figure*} [t!]
    \centering
    \includegraphics[width=0.99\textwidth]{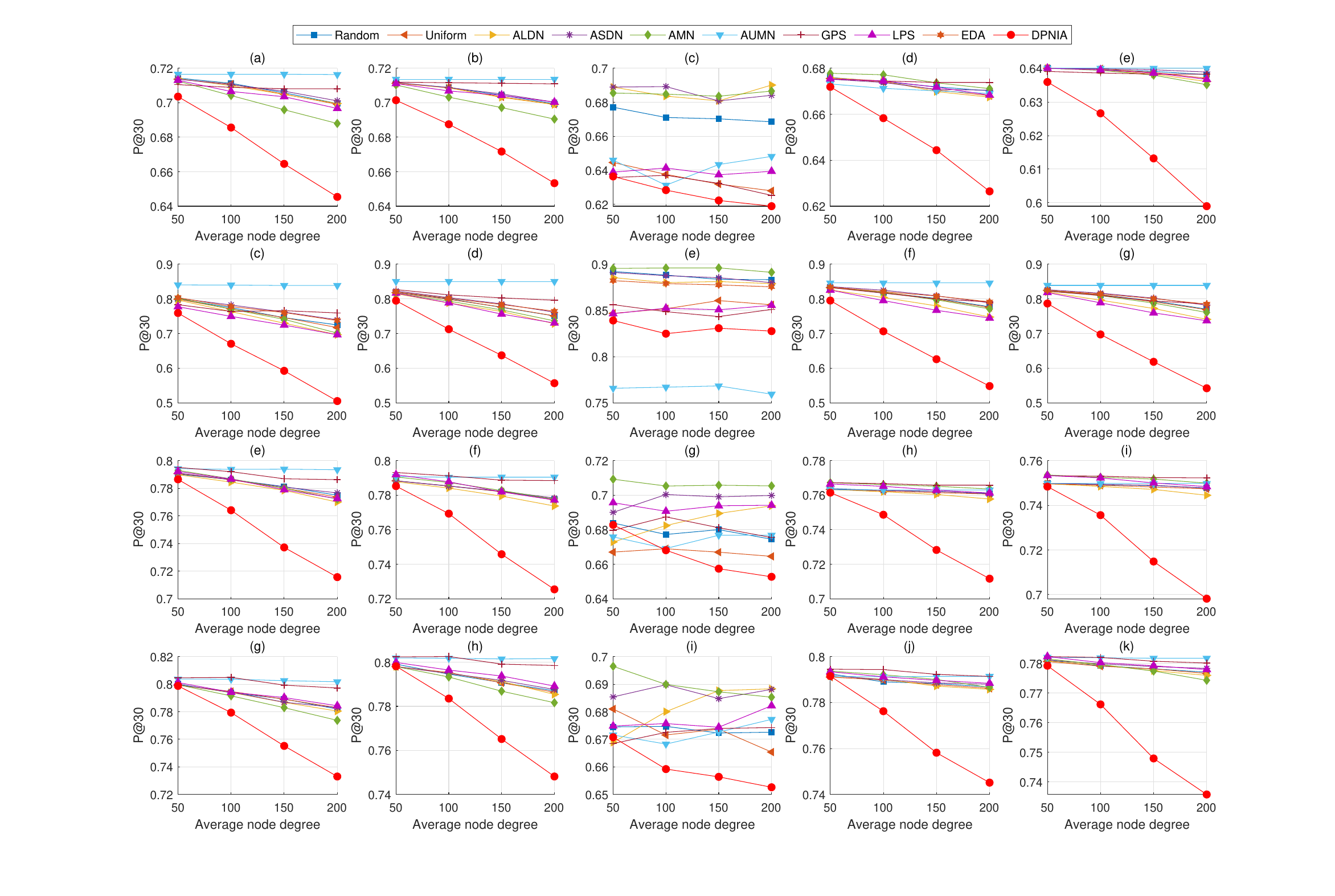}
    \caption{$P@30$ at different average node degree of injected nodes. The datasets and network alignment methods depicted in the various subfigures are consistent with those in Figure~\ref{pic_p30_diffNodeNum}.}
    \label{pic_p30_diffLinkNum}
\end{figure*}

Similarly, to achieve a comprehensive understanding of how various network alignment methods are affected by attacks on different datasets, we presented the details of the $P@30$ in Figure~\ref{pic_p30_diffLinkNum}. We can see that under the attack of DPNIA, almost all network alignment methods reached their lowest values of $P@30$ for a given average node degree. For instance, on the FT2 dataset when the average node degree is 200, after DPNIA's attack, the $P@30$ for SOIDP is $0.5053$, while the best-performing baseline, LPS, achieves $0.6962$. DPNIA results in a drop of $0.1909$. This observation once again underscores the effective performance of our proposed DPNIA in the context of network alignment. For the same network alignment method within a dataset, increasing the average node degree of injected nodes results in a smaller $P@30$ value. This aligns with our intuition that a larger attack budget leads to more substantial attacking performance. Apart from DPNIA, none of the attacking methods consistently outperformed the others across the majority of conditions. This, to some extent, once again provides evidence illustrating the effectiveness of our proposed method. Figure~\ref{pic_MAP_diffLinkNum} in Appendix A is the results of $MAP$ for varying the average node degree. We can see that the impact of varying the average node degree of injected nodes on the $MAP$ metric is quite consistent with that of $P@30$.

Based on the observations from Figure~\ref{pic_diffnodenum_avg} and Figure~\ref{pic_difflinknum_avg}, LPS demonstrates superior effectiveness in SNA attacks compared to all other baseline methods. To further substantiate the efficacy of our proposed DPNIA comprehensively, we conducted a detailed comparison of LPS and DPNIA under diverse conditions, encompassing injected node counts ranging from $100$ to $500$ and average degree of injected nodes varing from $50$ to $200$. Figure~\ref{pic_compLPSandDPNIA} presents the difference between the performance metrics corresponding to LPS and DPNIA. A larger difference indicates that, under those conditions, DPNIA outperforms LPS in enhancing network alignment attack effectiveness. From the figure, we can observe that DPNIA consistently outperforms LPS under all conditions. Specifically, when the number of injected nodes remains the same, DPNIA exhibits greater improvement in attack effectiveness over LPS as the average degree of injected nodes increases. Similarly, under the same average degree of injected nodes, DPNIA achieves more significant improvements in attack effectiveness as the number of injected nodes increases.

\subsection{Performance on different $@N$ settings}
\begin{figure} [t!]
    \centering
    \includegraphics[width=0.48\textwidth]{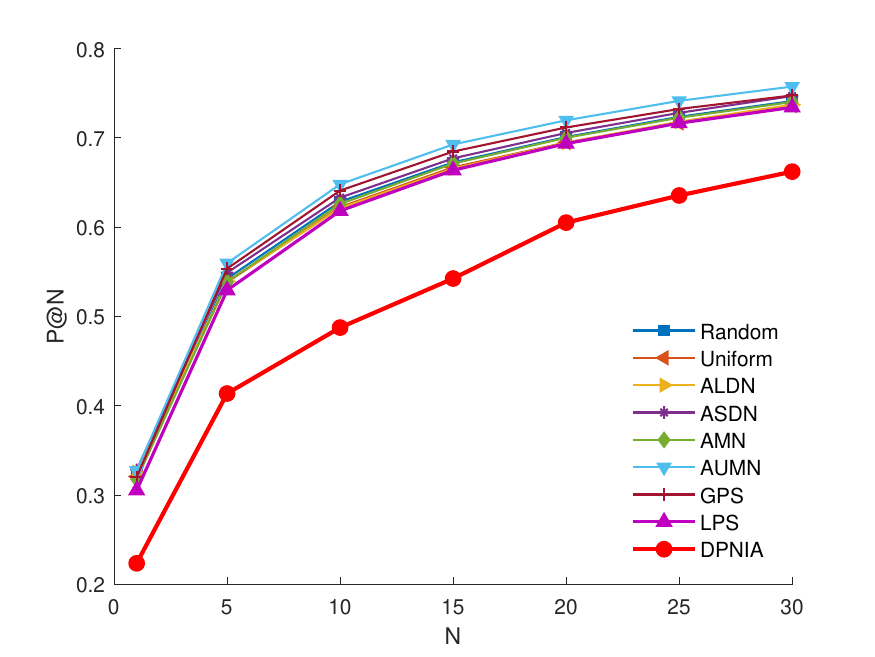}
    \caption{$P@N$ on different $N$.}
    \label{pic_diffPatN}
\end{figure}

We also evaluated the performance of the baselines and the proposed DPNIA at different $@N$ settings. We set the number of injected nodes to 200, with an average degree of 200, and performed the attack on a single network for each dataset. Figure~\ref{pic_diffPatN} displays the average $P@N$ of the five network alignment methods across various attacks applied to the four datasets. From the figure, we can see that DPNIA achieved the best attack performance, resulting in the lowest average values of $P@N$ for the five network alignment methods. For instance, under the DPNIA's attack, the average $P@N$ for the five network alignment methods decreased by an average of $0.0986$, with a maximum decrease of $0.1307$ compared to the best baseline attack method, LPS. Additionally, we can observe that as $N$ increases, the average $P@N$ values also increase. This is because $@N$ defines the length of the candidate pairing list for network alignment methods regarding an unmatched node. As $N$ increases, the candidate pairing list length grows, increasing the likelihood of correct pairings in the list, thus resulting in larger $P@N$ values. Nevertheless, regardless of the value of $N$, DPNIA consistently leads to the lowest average $P@N$ for network alignment methods. This reaffirms the effectiveness of our proposed DPNIA in node injection network alignment attacks.

\subsection{Evaluation via more metrics}
\begin{table*}[!t]
\centering
\caption{Assessment different attacking methods via more metrics. }
\setlength{\tabcolsep}{2.6pt}
\begin{tabular}{c|cccc|cccc|cccc|cccc}
\hline
{\multirow{2}*{}}&\multicolumn{4}{c|}{$\gamma(Precision)$}& \multicolumn{4}{c|}{$\gamma(Recall)$}& \multicolumn{4}{c|}{$\gamma(F1)$}& \multicolumn{4}{c}{$\gamma(AUC)$} \\
\cline{2-17}
~&FT1&TF2&FSMT&FSRT&FT1&TF2&FSMT&FSRT&FT1&TF2&FSMT&FSRT&FT1&TF2&FSMT&FSRT\\
\hline
Random&$0.948$&$0.815$&$0.932$&$0.956$&$0.995$&$0.936$&$0.979$&$1.012$&$0.951$&$0.824$&$0.936$&$0.959$&$0.905$&$0.890$&$0.905$&$0.902$\\
Uniform&$0.934$&$0.838$&$0.905$&$0.946$&$0.981$&$0.959$&$0.952$&$1.001$&$0.936$&$0.847$&$0.908$&$0.949$&$0.902$&$0.887$&$0.904$&$\uline{0.901}$\\
ALDN&$0.956$&$0.826$&$0.956$&$0.980$&$1.003$&$0.947$&$1.003$&$1.035$&$0.959$&$0.835$&$0.960$&$0.983$&$0.905$&$0.888$&$0.904$&$0.902$\\
ASDN&$0.964$&$0.868$&$0.954$&$0.979$&$1.011$&$0.989$&$1.001$&$1.034$&$0.967$&$0.877$&$0.957$&$0.982$&$0.906$&$0.910$&$0.905$&$0.903$\\
AMN&$0.950$&$0.817$&$0.923$&$0.901$&$0.997$&$0.933$&$0.974$&$0.956$&$0.953$&$0.825$&$0.927$&$0.905$&$0.902$&$0.883$&$0.907$&$0.903$\\
AUMN&$0.939$&$0.836$&$0.930$&$0.969$&$0.986$&$0.958$&$0.977$&$1.025$&$0.941$&$0.844$&$0.933$&$0.972$&$0.904$&$0.920$&$0.905$&$0.904$\\
GPS&$0.929$&$0.853$&$0.913$&$0.914$&$0.976$&$0.973$&$0.965$&$0.969$&$0.932$&$0.862$&$0.917$&$0.917$&$\uline{0.896}$&$0.882$&$\uline{0.902}$&$0.903$\\
LPS&$\uline{0.923}$&$\uline{0.730}$&$\uline{0.898}$&$\uline{0.889}$&$\uline{0.970}$&$\uline{0.849}$&$\uline{0.950}$&$\uline{0.944}$&$\uline{0.926}$&$\uline{0.739}$&$\uline{0.902}$&$\uline{0.893}$&$0.900$&$\uline{0.872}$&$0.903$&$0.904$\\
EDA&$-$&$0.846$&$-$&$-$&$-$&$0.961$&$-$&$-$&$-$&$0.854$&$-$&$-$&$-$&$0.893$&$-$&$-$\\
DPNIA&$\bm{0.582}$&$\bm{0.551}$&$\bm{0.636}$&$\bm{0.646}$&$\bm{0.624}$&$\bm{0.663}$&$\bm{0.681}$&$\bm{0.697}$&$\bm{0.582}$&$\bm{0.569}$&$\bm{0.639}$&$\bm{0.649}$&$\bm{0.888}$&$\bm{0.802}$&$\bm{0.888}$&$\bm{0.893}$\\
\hline
\end{tabular}
\label{tab_moremetrics}
\end{table*}
To comprehensively compare the effectiveness of different attack methods, we employed more metrics to assess their impacts for the network alignment in the subsequent experiments. The number of injected nodes in this group of experiments was set to 200, with an average degree of 200, and only one network within each dataset was subjected to the attack. We conducted an analysis of the ratios between each metric value during an attack and the corresponding metric value under non-attack conditions. The results are presented in Table~\ref{tab_moremetrics}. We can see that DPNIA exhibits the largest decline in all metrics. This once again underscores the effectiveness of our proposed method in this study. While LPS appears to be the second-best method in most cases, the gap between it and DPNIA remains quite significant. Apart from DPNIA and LPS, other baselines show limited disruption in SNA, with many metrics deviating by less than $5\%$ from their non-attacked values. An interesting phenomenon is the occurrence of $\gamma(Recall)$ values exceeding 1 for a few baselines. This suggests that under the respective method's attack, the network alignment's recall actually improves. The reasons are similar with those in Tables~\ref{tab:datasets_twolys} and Table~\ref{tab_datasets_lay2_200}.
\section{Conclusion}
In this study, we investigated the attack problem in SNA, aiming to perturb the structure of social networks in a practical and feasible manner to undermine the alignment algorithms or models that aligned account nodes belonging to the same user. We proposed a node injection attack framework based on dynamic programming (DPNIA) to achieve this goal, striving to ensure maximum interference with a minimal cost. This framework has the potential to safeguard against the misuse of SNA tools by malicious actors, thereby averting user privacy breaches, economic losses, and reputational damage. Meanwhile, it can be employed to evaluate the robustness of alignment algorithms and models, thus contributing to the promotion of beneficial applications for SNA tools. Although using the proposed DPNIA to perturb the social network structure can yield some level of effectiveness, this method is presently suitable for attacking only the task of social network alignment within the domain of network inference and primarily relied on human expertise and insights for conducting attacks. In the future, we plan to explore how to simultaneously generate attack effects on various network inference tasks and how to enable AI agents to find injection links that yield the best attack results.

\bibliographystyle{IEEEtran}
\bibliography{NodeInject_arXiv}
%


\vfill

\begin{appendices}
\textbf{Appendix}
\section{More results of the experiments}
\begin{figure*} [t!]
    \centering
    \includegraphics[width=0.99\textwidth]{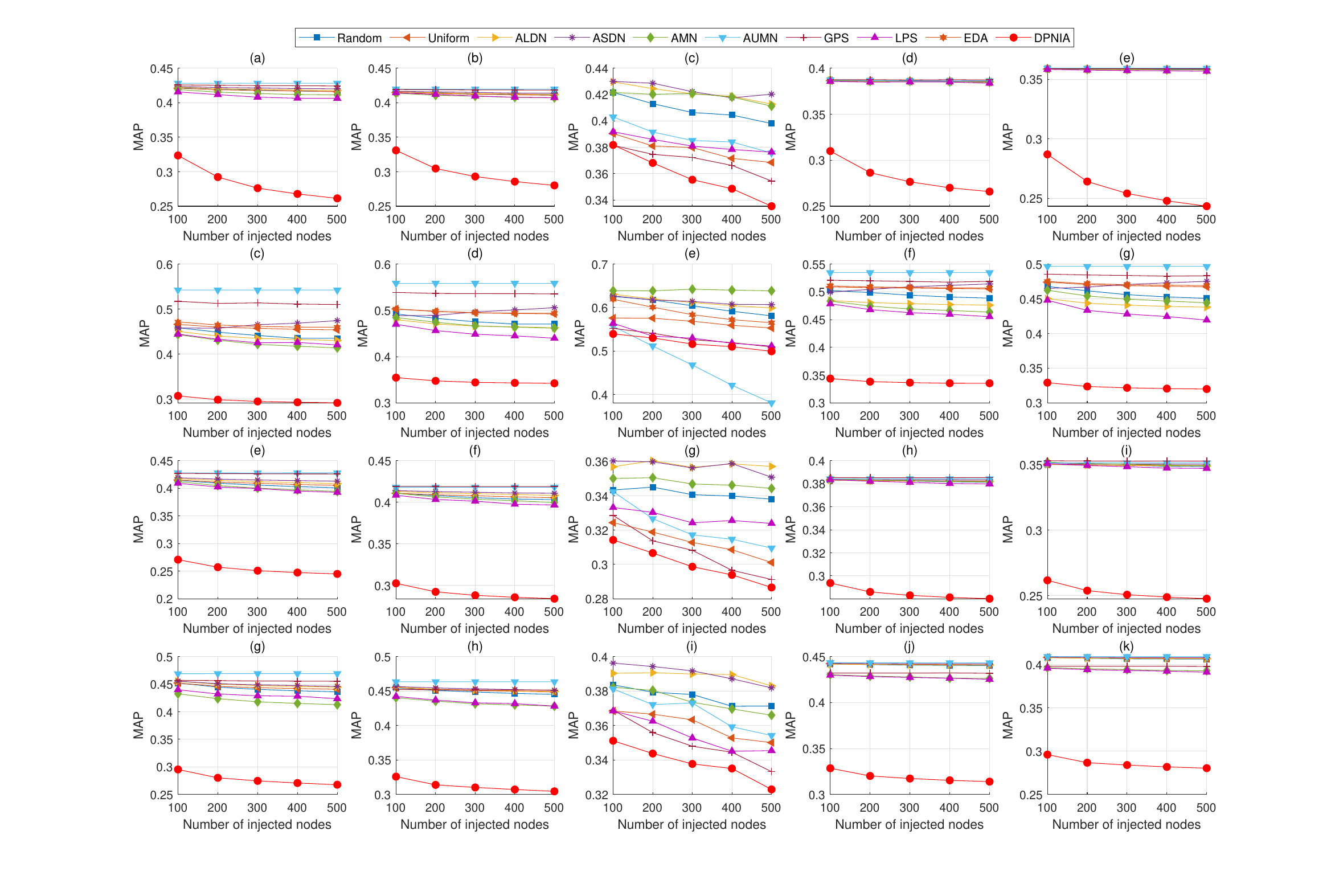}
    \caption{$MAP$ at different numbers of injected nodes. The datasets and network alignment methods depicted in the various subfigures are consistent with those in Figure~\ref{pic_p30_diffNodeNum}.}
    \label{pic_MAP_diffNodeNum}
\end{figure*}

Figure~\ref{pic_MAP_diffNodeNum} is the results of $MAP$ when varying the number of injected nodes. The trend of the curves in the figure closely resembles that in Figure~\ref{pic_p30_diffNodeNum}. Under the attack of DPNIA, almost all network alignment methods reached their lowest values of $MAP$ for a given number of injected nodes. For instance, on the FSRT dataset when the number of injected node is 100, after DPNIA's attack, the $MAP$ for CN is $0.2954$, while the best-performing baseline, LPS, achieves $0.4325$. DPNIA results in a drop of $0.1371$. For the same network alignment method within a dataset, increasing the average node degree of injected nodes results in a smaller $MAP$ value. In addition, apart from DPNIA, none of the attacking methods consistently outperformed the others across the majority of conditions.

\begin{figure*} [t!]
    \centering
    \includegraphics[width=0.99\textwidth]{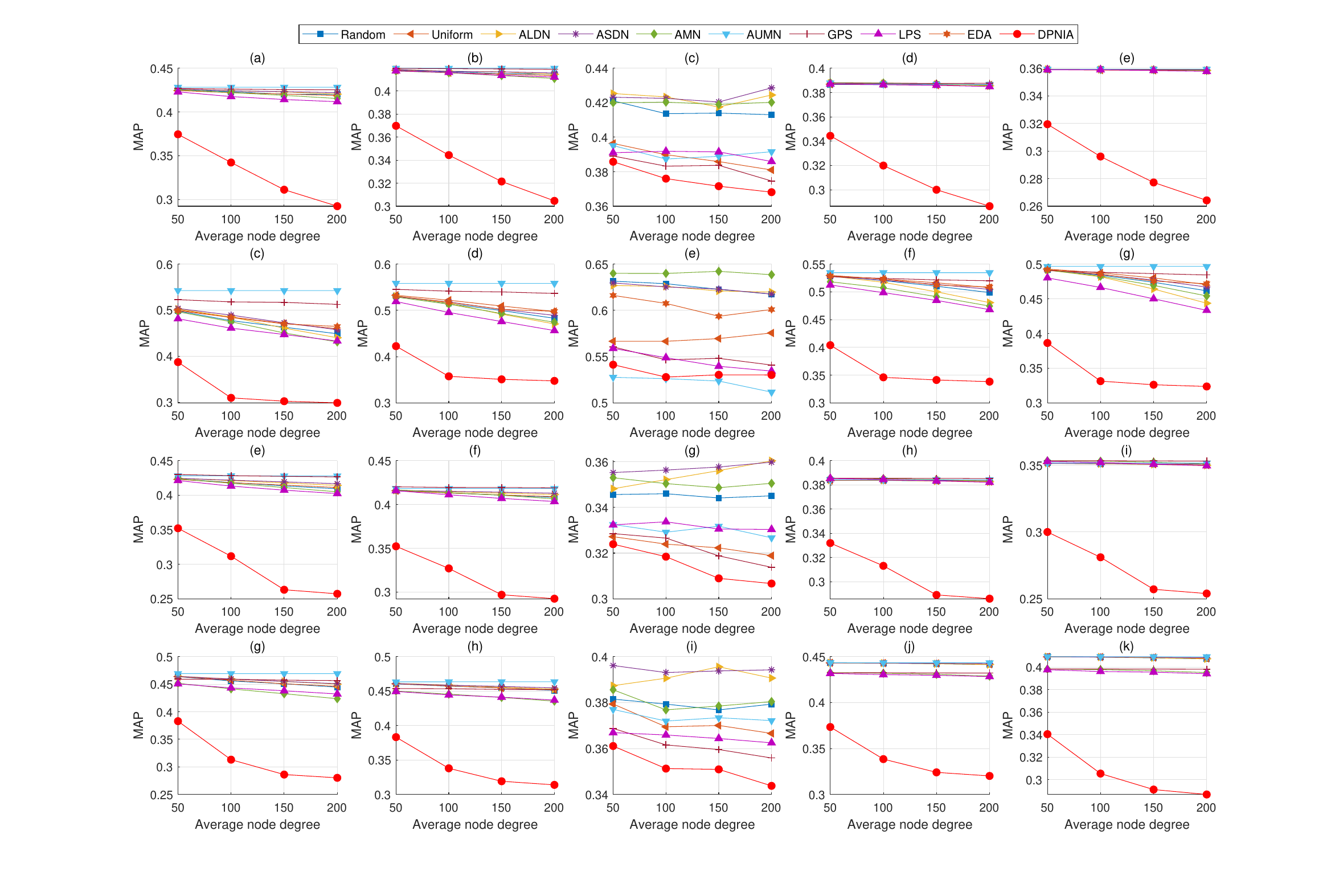}
    \caption{$MAP$ at different average node degree of injected nodes. The datasets and network alignment methods depicted in the various subfigures are consistent with those in Figure~\ref{pic_p30_diffNodeNum}.}
    \label{pic_MAP_diffLinkNum}
\end{figure*}

Figure~\ref{pic_MAP_diffLinkNum} is the results of $MAP$ when varying the average node degree of the injected nodes. The trend of the curves in the figure closely resembles that in Figure~\ref{pic_p30_diffLinkNum}. Under the attack of DPNIA, almost all network alignment methods reached their lowest values of $MAP$ for a given average node degree. For instance, on the FSMT dataset when the average node degree is 200, after DPNIA's attack, the $MAP$ for SOIDP is $0.2802$, while the best-performing baseline, LPS, achieves $0.4238$. DPNIA results in a drop of $0.1436$. For the same network alignment method within a dataset, increasing the average node degree of injected nodes results in a smaller $MAP$ value. In addition, apart from DPNIA, none of the attacking methods consistently outperformed the others across the majority of conditions.
\end{appendices}

\end{document}